\documentclass[pdflatex,sn-mathphys-num]{sn-jnl}

\usepackage{graphicx}%
\usepackage{multirow}%
\usepackage{amsmath,amssymb,amsfonts}%
\usepackage{amsthm}%
\usepackage{mathrsfs}%
\usepackage[title]{appendix}%
\usepackage{xcolor}%
\usepackage{textcomp}%
\usepackage{manyfoot}%
\usepackage{booktabs}%
\usepackage{algorithm}%
\usepackage{algorithmicx}%
\usepackage{algpseudocode}%
\usepackage{listings}%
\usepackage{graphicx}
\usepackage{caption}
\usepackage{subcaption}
\usepackage{overpic}

\theoremstyle{thmstyleone}%
\theoremstyle{thmstyletwo}%

\theoremstyle{thmstylethree}%

\begin{document}

%title[Article Title]{Bi-TEAM: Unifying Biochemical Representations for Non-Canonical Peptide Property Prediction}
\title[Article Title]{Bi-TEAM: A Unified Cross-Scale Representation Learning Framework for Chemically Modified Biomolecules}


\author[1]{\fnm{Chunbin} \sur{Gu}}
\equalcont{These authors contributed equally to this work.}

\author[1]{\fnm{Zijun} \sur{Gao}}
\equalcont{These authors contributed equally to this work.}

\author[4]{\fnm{Mutian} \sur{He}}

\author[1]{\fnm{Jingjie} \sur{Zhang}}

\author[5]{\fnm{Haipeng} \sur{Wen}}

\author[6]{\fnm{Zihao} \sur{Luo}}

\author[2]{\fnm{Xiaorui} \sur{Wang}}

\author*[1]{\fnm{Hanqun}
\sur{Cao}}\email{hanquncao2001@link.cuhk.edu.hk}

\author*[3]{\fnm{Jiajun}
\sur{Bu}}\email{bjj@zju.edu.cn}

\author*[2]{\fnm{Chang-Yu} \sur{Hsieh}}\email{kimhsieh@zju.edu.cn}

\author[1]{\fnm{Pheng Ann} \sur{Heng}}

\affil[1]{\orgdiv{Department of Computer Science and Engineering}, \orgname{The Chinese University of Hong Kong}, \orgaddress{\state{Hong Kong}}}

\affil[2]{\orgdiv{College of Pharmaceutical Sciences}, \orgname{Zhejiang University}, \orgaddress{\city{Hangzhou}, \postcode{310027}, \state{Zhejiang Province}, \country{China}}}

\affil[3]{\orgdiv{College of Computer Science and Technology}, \orgname{Zhejiang University}, \orgaddress{\city{Hangzhou}, \postcode{310007}, \state{Zhejiang Province}, \country{China}}}

\affil[4]{\orgdiv{Faculty of Applied Sciences}, \orgname{Macao Polytechnic University}, \orgaddress{\state{Macao}}}

\affil[5]{\orgdiv{Department of Ophthalmology}, \orgname{The Second Xiangya Hospital of Central South University}, \city{Changsha}, \postcode{410011}, \state{Hunan Province}, \country{China}}

\affil[6]{\orgdiv{School of Mechanical and Electrical Engineering}, \orgname{University of Electronic Science and Technology of China}, \orgaddress{\city{Chengdu}, \postcode{611731}, \state{Sichuan Province}, \country{China}}}


\abstract{
Representation learning for protein biochemical space faces a difficult trade-off: protein language models excel at capturing long-range biological semantics but often miss fine-grained chemical details. Conversely, chemical language models encode atomic information but lack the broader sequence context. To address this, we introduce Bi-TEAM (Bi-gated Residual Space Modification), a general framework that injects localized chemical variation into global protein contexts. By ensuring robustness against perturbations like non-canonical amino acids, post-translational modifications (PTMs), and topological constraints, Bi-TEAM uncovers functional chemical dependencies often missed by evolutionary baselines. Mechanistically, Bi-TEAM maps non-canonical residues to their natural counterparts and injects atomic-level data via a bi-gated residual fusion mechanism. Crucially, this process uses modification-aware prompts to ensure that local structural changes influence global functional representations—all without the need for alphabet expansion. We tested Bi-TEAM on ten datasets spanning application-critical chemically modified peptides, PTMs, and natural proteins. The model consistently outperformed state-of-the-art baselines, achieving up to a 66\% improvement in Matthews correlation coefficient (MCC) on rigorous scaffold-similarity splits and a 350\% boost in hemolysis prediction accuracy. Furthermore, when deployed as an oracle for generative modeling, Bi-TEAM nearly quadrupled the success rate of designing cell-penetrating cyclic peptides. By unifying biological semantics with chemical precision, Bi-TEAM offers a versatile foundation for the machine learning-driven exploration of peptide and protein biochemical space.}


\maketitle

\section{Introduction}\label{intro}

Representation learning has emerged as a pivotal paradigm in biochemistry~\cite{gawehn2016deep, butler2018machine}, where the quality of learned embeddings fundamentally dictates performance across downstream tasks~\cite{rives2021biological, wang2022molecular}. However, as single-modality frameworks, the prevalent Protein Language Models (PLMs) and Chemical Language Models (CLMs) often fail to capture the intricate protein chemical matter essential for function~\cite{paton2025connecting,zhang2025learning}, thereby segregating biological evolution from chemical rationality. This limitation is particularly detrimental when navigating the expanded biochemical space where non-canonical amino acids (ncAAs) introduce complex chemical modifications. While such modifications transcend the functional boundaries of natural polypeptides to confer therapeutic advantages like enhanced stability and bioavailability \cite{li2024therapeutic, zorzi2017cyclic, oeller2023sequence}, they impose significant modeling challenges.

Given that such chemical versatility is most effectively realized in peptide engineering, peptides represent a prime exemplar of this hybrid chemical-biological landscape~\cite{salveson2024expansive}. In sequence-based peptide modeling, Protein Language Models such as ESM, ProtT5, and PeptideBERT have reshaped the field by treating amino acid sequences as a “language of life” \cite{lin2023evolutionary, schmirler2024fine, guntuboina2023peptidebert}. By learning from large-scale sequence corpora, these models capture contextual and evolutionary regularities that often correlate with peptide structure and function, providing transferable representations for downstream prediction and design tasks. Recent advances—including PepPrCLIP, LassoESM, and PepMLM—have further adapted PLM-derived representations to peptide-specific applications via task-tailored objectives and architectures \cite{bhat2025novo, mi2024lassoesm, chen2024pepmlm, brixi2023salt}.

However, the incorporation of ncAAs presents a critical challenge for standard PLMs restricted to the canonical alphabet. These models inherently struggle to tokenize chemically modified residues. Approximation strategies like PepFun substitute ncAAs with natural counterparts, yet this introduces significant bias by ignoring precise chemical alterations \cite{ochoa2025pepfunn}. Conversely, vocabulary expansion approaches such as MuCoCP and GPepT theoretically capture diversity but suffer from sparse semantic density due to an unmanageable proliferation of tokens \cite{yu2024mucocp, oikawa2025gpept}.

To overcome these limitations, researchers have shifted to a chemical perspective, treating peptides as precise atomic assemblies rather than mere sequences of letters. To capture this atomistic resolution, Chemical Language Models have been developed to represent peptides using fixed chemical tokens, a strategy that inherently accommodates non-canonical amino acids. While generic CLMs trained on small molecules often suffer from domain gaps \cite{chithrananda2020chemberta, irwin2022chemformer, edwards2022translation}, peptide-specific adaptations, such as PeptideCLM and PepDoRA, have demonstrated potential in predicting the properties of modified peptides \cite{feller2024peptide, wang2024pepdora}. Nevertheless, relying strictly on chemical representations can obscure the global biological context. Moreover, the resulting expansion of peptide sequences into dense chemical tokens often exceeds the context windows of current models, thereby limiting their applicability to longer sequences.

We posit that chemical variation in peptides should be modeled as structured, localized perturbations to an underlying biological semantic space, rather than as an independent representation to be jointly learned. This motivates a selective fusion paradigm in which biological context remains dominant while chemical information is injected where functionally relevant. To this end, we introduce Bi-gaTed rEsidual spAce Modification (Bi-TEAM), a framework that adaptively integrates multi-scale biochemical properties to serve as both a robust predictor and a high-fidelity oracle for efficient peptide design.

Specifically, Bi-TEAM constructs a dual-view representation that harmonizes evolutionary biological space with fine-grained chemical space. We map ncAAs to their natural counterparts for PLM encoding to preserve global evolutionary semantics, while simultaneously employing a CLM to capture atomistic molecular details. To integrate these modalities, we devise a bi-gated residual mechanism—guided by a position-aware modification prompt that indicates where an ncAA occurs—that treats the biological representation as a semantic backbone and adaptively injects chemical signals as precise modifications. This design ensures that the surrounding biochemical context is dynamically adapted to represent modified residues, thereby explicitly linking local chemical alterations to macroscopic functional effects while balancing global sequence constraints with local physicochemical interactions.

We conducted a comprehensive evaluation across three biochemical domains spanning ten diverse datasets, where Bi-TEAM established new state-of-the-art benchmarks on seven critical predictive tasks. These tasks encompass membrane permeability for modified peptides~\cite{geylan2024methodology,bhardwaj2022accurate, yu2024mucocp, li2023cycpeptmpdb, rezai2006conformational}, druggability and pathogenicity for post-translational modifications (PTMs)\cite{peng2024ptm,peng2025ptm}, and cell-penetrating efficiency\cite{kumar2025plm4cpps}, alongside physicochemical properties such as hemolysis, solubility, and non-fouling for natural peptides~\cite{ansari2023serverless, smialowski2012proso, barrett2018classifying}.

Bi-TEAM demonstrates exceptional generalization across three biochemical domains. First, for non-canonical peptides, it improves MCC by over 15\% on random splits and surpasses state-of-the-art methods by 66\% on challenging scaffold-similarity splits. Second, for PTMs, it increases druggability prediction accuracy by 18\%. Third, regarding natural peptides and proteins, it achieves a 350\% MCC gain in hemolysis prediction. Beyond prediction, when deployed as a generative oracle, Bi-TEAM boosts the yield of valid cell-penetrating cyclic peptides by nearly four-fold, validating its utility in de novo design.

Finally, interpretability analyses confirm that Bi-TEAM effectively captures and quantifies the impact of critical chemical modifications—such as N-methylation and cis-peptide bond formation—on transmembrane permeability, aligning well with medicinal chemistry intuition. By bridging the gap between evolutionary biological semantics and fine-grained chemical representations, Bi-TEAM serves as a robust, extensible foundation for generalized peptide design, paving the way for accelerated therapeutic discovery.

\begin{figure}[htbp]
    \centering
        \includegraphics[width=\textwidth]{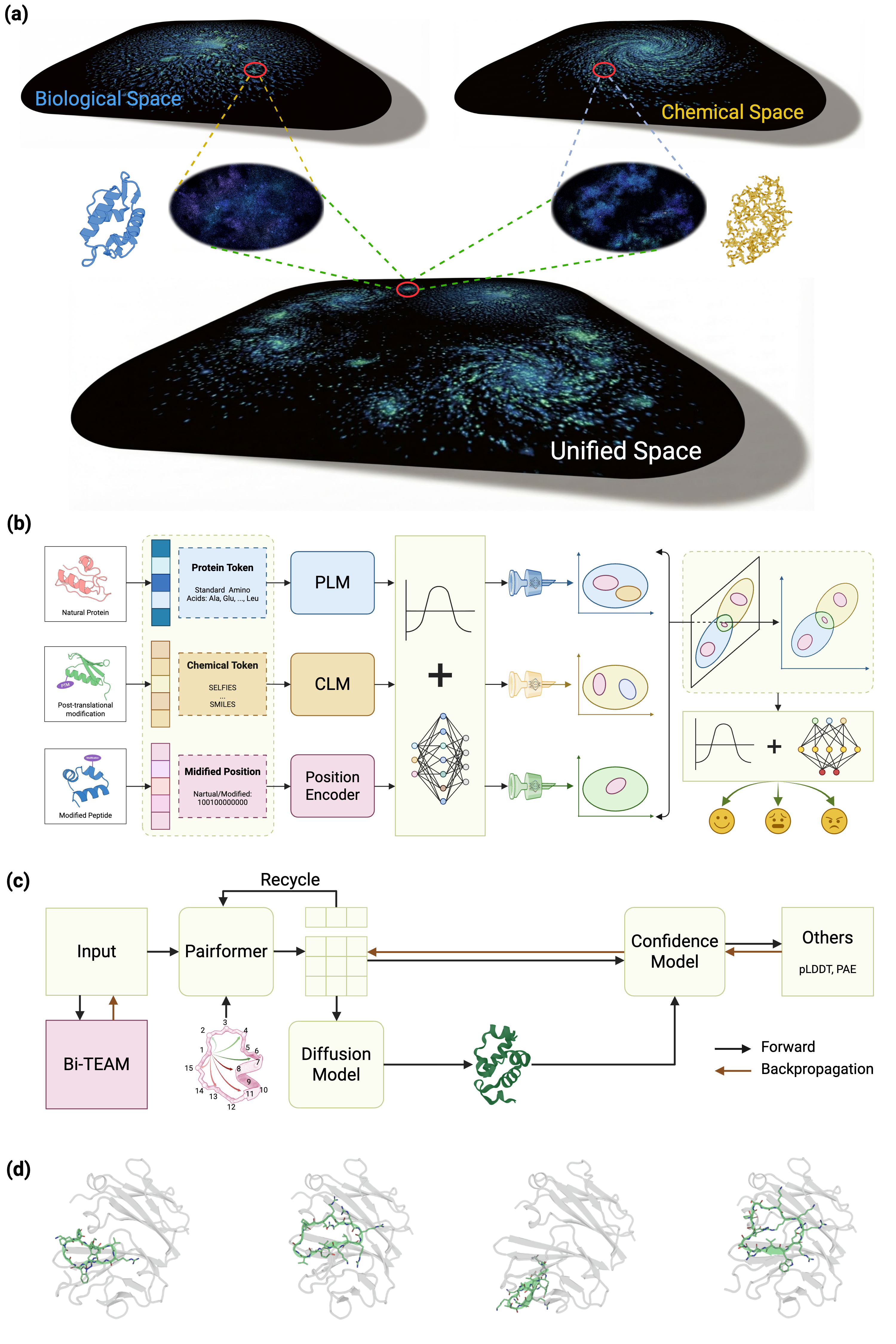}
    \caption{\textbf{Overview of the Bi-TEAM framework.} 
    \textbf{(a)}~Exploration of a unified protein representation space by integrating biological-based and chemical-based feature spaces. 
    \textbf{(b)}~Network architecture of Bi-TEAM focusing on multi-property prediction tasks, which comprehensively utilizes multi-modal information from the Protein Language Module, Chemical Language Module, and Modified Position module, exploring the optimal space by fusing these different modalities.
    \textbf{(c)}~Conditional generation of modified peptides with specific properties, achieved by using Bi-TEAM to guide the BoltzDesign1. 
    \textbf{(d)}~Case studies illustrating generated modified peptides which possess superior membrane permeability.}
    \label{framework}
\end{figure}

\section{Results}
\subsection{Overview of Bi-TEAM} Bi-gaTed rEsidual spAce Modification (Bi-TEAM) establishes the biological space constructed by Protein Language Models (PLMs) as its foundation to fully leverage broad biological context, while strategically employing Chemical Language Models (CLMs) to capture fine-grained chemical details and further explore the boundaries of the input space. Specifically, each modification-containing input is processed through two distinct streams: (i) a biological sequence in which each modified amino acid is mapped to its closest natural amino acid —thereby mitigating token inflation while preserving valuable evolutionary cues for the PLM—and (ii) a SELFIES-like representation that encodes atomic-level modifications for the CLM.

After encoding both streams, Bi-TEAM fuses amino acid-level embeddings with atom-level embeddings, effectively balancing the intrinsic strengths of each domain. Additionally, a specifically designed ``Modified Position" indicator pinpoints where modifications appear, allowing the network to focus on key sites of chemical variation. Notably, the model also demonstrates versatility by accommodating datasets without modification by omitting both the token mapping and modification location steps.

To validate the generalizability of the proposed framework, we conducted tests across multiple properties in three distinct domains: modified peptides, post-translational modifications (PTMs), and natural peptides and proteins. For modified peptides, the focus is primarily on membrane permeability; for PTMs, the focus is on druggability and disease associations; and for natural peptides and proteins, the focus is on hemolysis prediction, solubility assessment, and nonfouling properties. Each dataset addresses a distinct aspect of therapeutic molecule functionality that directly impacts clinical viability, from safety concerns to delivery capabilities, providing a comprehensive framework for evaluating computational methods across the drug discovery pipeline. As a result, Bi-TEAM demonstrates exceptional performance across all tasks, offering a scalable and robust solution for modeling across diverse domains.

\subsection{Contrastive Evaluations Across Multiple Domains}
\begin{figure}[htbp]
    \centering
        \includegraphics[width=\textwidth]{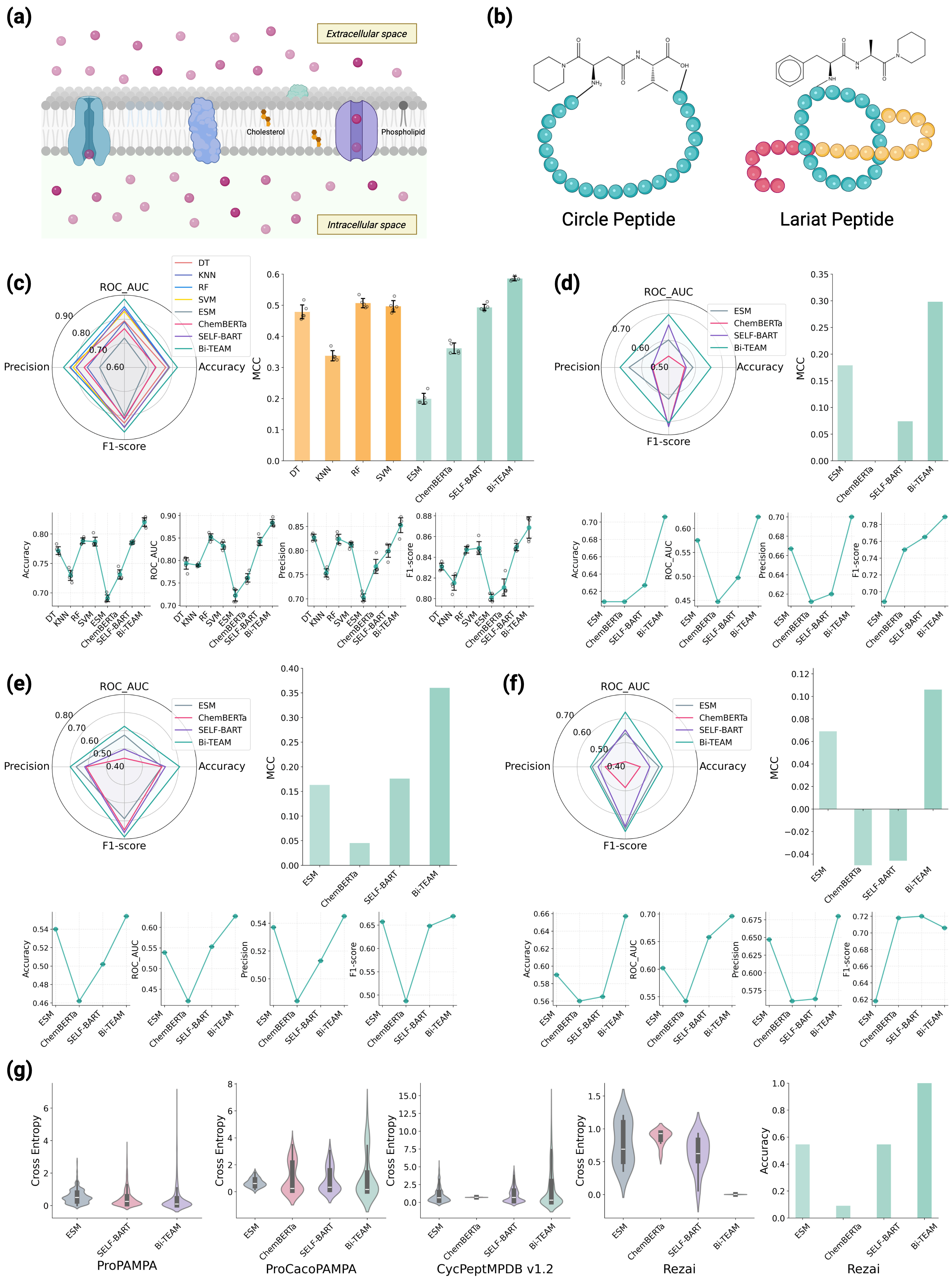}
\caption{\textbf{Comparative analysis of model performance on different modified peptide datasets for permeability prediction.} \textbf{(a)} Schematic illustration of the principle of cell membrane permeability. \textbf{(b)} Representative examples of modified peptides, including cyclic peptides and lariat peptides. \textbf{(c)} Performance evaluation on the modified ProPAMPA dataset \cite{li2023cycpeptmpdb, geylan2024methodology}. \textbf{(d)} Generalization assessment using a principled cluster-based train-test partitioning strategy derived from fingerprint similarity distributions within the modified ProPAMPA dataset. \textbf{(e–f)} Inference results on the modified ProCocaPAMPA dataset \cite{bhardwaj2022accurate,yu2024mucocp} and CycPeptMPDB v1.2 \cite{li2023cycpeptmpdb}, respectively, using models pre-trained on ProPAMPA. \textbf{(g)} Cross-entropy loss for membrane permeability prediction on ProPAMPA, ProCacoPAMPA, and Rezai datasets \cite{rezai2006conformational} via direct inference using ProPAMPA-trained models, alongside accuracy metrics for the Rezai dataset. Radar charts display the average values across four metrics, while bar charts present detailed individual results with discrete data points; error bars indicate standard deviation. (Results for \textbf{(d)}, \textbf{(e)}, and \textbf{(f)} are identical across the five runs as they represent direct inference.)}
\label{ModifiedPeptide}
\end{figure}

\begin{figure}[htbp!]
    \centering
\includegraphics[width=1.0\textwidth]{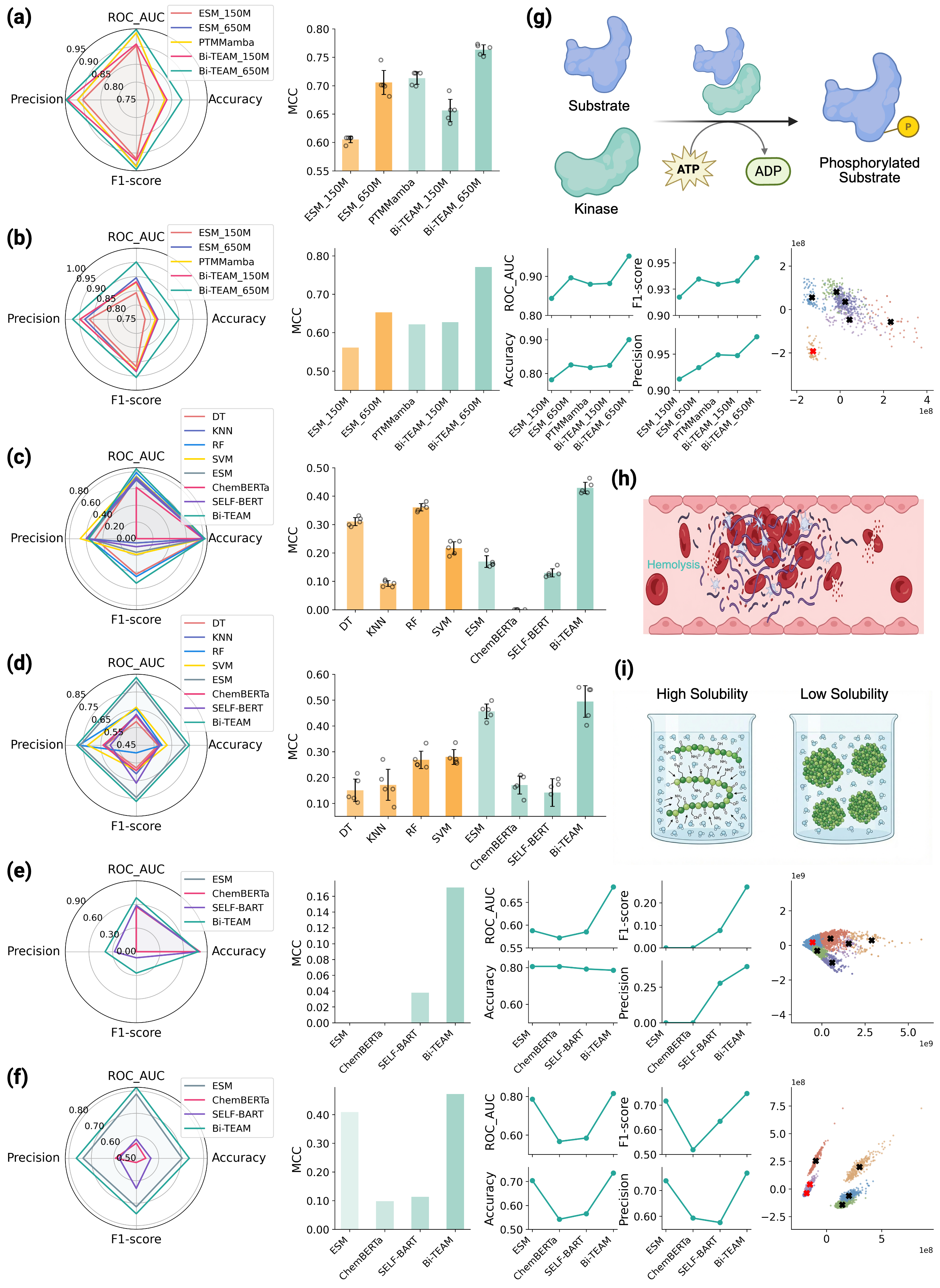} 
    \caption{\textbf{Generalization assessment on PTM and natural protein datasets.} \textbf{(a, b)} Predictive performance evaluation for PTM druggability across five metrics under random splitting \textbf{(a)} and similarity-based splitting \textbf{(b)} strategies. \textbf{(c, d)} Comparative prediction results for natural protein hemolysis \textbf{(c)} and  solubility \textbf{(d)} using random data splits. \textbf{(e, f)} Corresponding performance assessments for hemolysis \textbf{(e)} and solubility \textbf{(f)} under similarity-based splits. For panels \textbf{(a)}, \textbf{(c)}, and \textbf{(d)}, the mean, standard deviation, and specific values for five random runs are reported. The rightmost columns in \textbf{(b)}, \textbf{(e)}, and \textbf{(f)} present t-SNE visualizations of the data distributions under similarity-based splitting, where black and red crosses designate the training and test sets, respectively. \textbf{(g--i)} Schematic illustrations depicting the underlying mechanisms of PTM \textbf{(g)}, hemolysis \textbf{(h)}, and solubility \textbf{(i)}.}
    \label{NaturalPTM}
\end{figure}

In this study, we conducted a comprehensive evaluation of both property prediction and property-guided generation. For the prediction tasks, we utilized 10 datasets spanning three domains: modified peptides (Figure~\ref{ModifiedPeptide}), post-translational modifications and natural proteins (Figure~\ref{NaturalPTM}). Prediction performance was assessed using metrics including the Matthews Correlation Coefficient (MCC) \cite{matthews1975comparison,baldi2000assessing}, Accuracy, ROC-AUC, Precision, and F1-score (see Appendix \ref{metric} for detailed descriptions). For the generation tasks, our primary focus was designing cell-penetrating cyclic peptides; therefore, the evaluation metrics centered on cell-penetrating capability.

\subsubsection{Benchmarking Modified Peptides: Fusing Bio-Syntax with Chem-Topology}

We first focused on membrane permeability prediction for modified peptides by training our model on ProPAMPA—a publicly accessible database for modified cyclic/lasso peptide membrane permeability \cite{geylan2024methodology}. We compared its performance against four classical machine learning algorithms (Decision Tree (DT) \cite{myles2004introduction}, K-Nearest Neighbors (KNN) \cite{cover1967nearest}, Random Forest (RF) \cite{breiman2001random}, and Support Vector Machine (SVM) \cite{cortes1995support}) and three deep learning approaches (ESM \cite{lin2023evolutionary}, ChemBERTa \cite{chithrananda2020chemberta}, and SELF-BART \cite{priyadarsini2024self}). 

To validate generalization capability, we evaluated the ProPAMPA-trained deep learning model on three external, wet-lab validated datasets—ProCacoPAMPA \cite{bhardwaj2022accurate}, CycPeptMPDB v1.2 \cite{li2023cycpeptmpdb}, and Rezai \cite{rezai2006conformational}—demonstrating its potential as an effective tool for drug candidate screening.

Figure \ref{ModifiedPeptide}(a) illustrates the fundamental mechanisms of membrane permeability. Driven by concentration and electrochemical gradients, molecules traverse the cellular membrane via selective pathways—either simple diffusion or protein-mediated transport—governed by the interplay between their physicochemical properties, the lipid bilayer, and membrane proteins. To capture the structural diversity of therapeutic candidates, Figure \ref{ModifiedPeptide}(b) depicts representative modified peptides, including cyclic and lasso peptides, which are central to our study.

\textbf{Performance on the modified ProPAMPA dataset (random split).} We employ multiple metrics to analyze the experimental results, with MCC being particularly suitable for performance evaluation given ProPAMPA’s imbalanced 2:1 positive-to-negative ratio. Our proposed framework, Bi-TEAM, demonstrates superior performance, achieving an MCC nearly 2.5 times higher than the protein language model ESM, which represents the biological subspace. Bi-TEAM also secures substantial gains across other metrics, with approximate improvements of +18.6\% in Accuracy, +22.4\% in ROC-AUC, +21.3\% in Precision, and +8.3\% in F1-score. Since modified amino acids often involve subtle yet critical chemical alterations, ESM's reliance on natural amino acid representations inherently limits its ability to capture the impact of these modifications on permeability. By integrating explicit chemical insights, Bi-TEAM effectively mitigates the limitations associated with modeling based purely on amino acid sequences.

The chemical encoder SELF-BART utilized within our framework outperforms ESM but remains subordinate to Bi-TEAM across multiple metrics. Specifically, Bi-TEAM delivers relative improvements of approximately +18.9\%, +4.3\%, +5\%, +6.7\%, and +2\% in MCC, Accuracy, ROC-AUC, Precision, and F1-score, respectively, compared to SELF-BART. This disparity indicates that purely chemical modeling may overlook essential peptide-level biological context, reinforcing the necessity of integrating both sequence-level biological cues and atom-functional-group–level chemical information. A comparison between PLM and CLM frameworks reveals that CLM-based approaches, such as SELF-BART, inherently handle chemical modifications more effectively, supporting the premise that even modest chemical modifications can significantly modulate a peptide's membrane permeability. Bi-TEAM effectively unifies these complementary representations to achieve superior predictive performance. Furthermore, when comparing the SELFIES-based SELF-BART with the SMILES-based ChemBERTa, we observed that SELF-BART consistently yields higher Accuracy and F1-scores by several percentage points. This advantage is attributed to the robust and syntactically unambiguous nature of the SELFIES representation, which minimizes invalid molecular expressions and provides a more consistent foundation for downstream learning tasks. 

While deep learning architectures like ChemBERTa and SELF-BART offer richer representations for large or highly diverse datasets, classical machine learning models—particularly SVM and Random Forest—remain highly competitive for small-to-medium-sized datasets such as ProPAMPA. These traditional methods typically rely on SMILES-based fingerprints, which yield interpretable chemical features and exhibit a lower risk of overfitting. With carefully curated feature engineering, these models can occasionally match or even exceed neural networks in data-limited scenarios. Figure~\ref{ModifiedPeptide}(c) summarizes the comparative experimental results across all deep learning and traditional machine learning methods. As illustrated in the MCC bar chart, Bi-TEAM achieves an approximate +15\% improvement over the second-best baseline. The radar and bar charts further corroborate Bi-TEAM's consistent superiority across other metrics, showing relative improvements of approximately +4\% in Accuracy, +3.8\% in ROC-AUC, +3\% in Precision, and +2\% in F1-score. Furthermore, we provide a granular performance analysis using confusion matrices in Appendix~\ref{ConfusionMatrix} to elucidate the model's classification behavior.

\textbf{Generalization to Novel Scaffolds (Similarity-Based Split).} To evaluate the model's generalization capability regarding structurally novel peptides and to better reflect real-world application scenarios involving predictions on unseen peptide families, we implemented similarity-based splits alongside traditional random splits. Utilizing a fingerprint similarity-based data splitting strategy (detailed in Appendix \ref{dataSplitMethods}), we conducted a comprehensive evaluation on the modified ProPAMPA dataset, with results shown in Figure~\ref{ModifiedPeptide}(d). Overall, compared to random partitioning, all models experienced performance degradation across various metrics due to the distributional shift. Nevertheless, Bi-TEAM demonstrated significant resilience compared to other models, highlighting its robustness under stringent data partitioning schemes. Specifically, Bi-TEAM further expanded its lead over the best baseline in terms of MCC, achieving a 66.67\% improvement. Notably, the stringent data splitting triggered a precipitous drop in ChemBERTa's MCC to 0. Bi-TEAM also achieved substantial relative gains of approximately 11.36\%, 5.78\%, and 5.10\% in Accuracy, ROC-AUC, and Precision, respectively, while maintaining a highly competitive F1-score—slightly lower than ChemBERTa and SELF-BART but still excellent.

\textbf{Validation on External Wet-Lab Datasets.} While similarity-based splits simulate real-world scenarios, direct evaluation on wet-lab validated external test sets offers more persuasive evidence of generalizability. To explore our model's potential as an early-stage drug screening tool, we assessed its performance on three wet-lab validated external test sets distinct from ProPAMPA, which serve as reliable benchmarks for the practical utility of our computational method. These test sets encompass molecules with diverse scaffolds and novel structural modifications, simulating the varied molecular space typical of later-stage drug discovery. We utilized t-SNE \cite{van2008visualizing} to reduce the 2048-dimensional features to two dimensions for visualization, as shown in Appendix~\ref{ood_wet}. Figures~\ref{ModifiedPeptide}(e) and (f) illustrate the inference performance of Bi-TEAM, trained on ProPAMPA, when applied to the external datasets ProCacoPAMPA~\cite{bhardwaj2022accurate} and CycPeptMPDB v1.2~\cite{li2023cycpeptmpdb}. The radar and MCC bar charts reveal that our approach consistently outperforms baseline methods by a considerable margin across five core metrics on these datasets. These experiments demonstrate that single-domain models (e.g., ESM, ChemBERTa, SELF-BART) may emphasize specific advantages but fail to excel comprehensively across all metrics.

Compared to CLMs, ESM exhibits advantages in most metrics on both external datasets. In contrast to its weaker performance on ProPAMPA, ESM displays excellent generalization capabilities on external data, even without explicit training on modifications. However, it still fails to capture specific modification information during inference. By utilizing ESM to construct the foundational biological peptide space, Bi-TEAM demonstrates significantly enhanced performance once the key chemical modifications are incorporated. On the ProCacoPAMPA and CycPeptMPDB v1.2 datasets, Bi-TEAM increases MCC by 2.2 times and 1.5 times relative to ESM, respectively. Additionally, Accuracy, F1-Score, ROC-AUC, and Precision improve by 16.1\%, 14.7\%, 8.5\%, and 4.9\% on ProCacoPAMPA, and by 2.6\%, 1.9\%, 16.3\%, and 1.5\% on CycPeptMPDB v1.2. Regarding CLMs, a domain discrepancy exists between the small-molecule datasets used in pretraining and the peptide-based molecular data encountered during inference, resulting in performance inferior to ESM. SELF-BART benefits from larger training data and stronger SELFIES representations compared to ChemBERTa, and Bi-TEAM further amplifies this advantage. Our comparative analysis reveals that Bi-TEAM achieves a remarkable 105\% increase in MCC on the ProCacoPAMPA dataset over SELF-BART. Notably, on the CycPeptMPDB v1.2 dataset, where SELF-BART exhibits negative performance (MCC = $-0.046$), Bi-TEAM maintains robustness with an MCC of $0.106$. This superiority extends across multiple metrics, with Bi-TEAM improving Accuracy, F1-Score, ROC-AUC, and Precision by 12.6\%, 3.1\%, 25.6\%, and 12.9\% on ProCacoPAMPA, and 10.4\%, 3.2\%, 13.4\%, and 6.2\% on CycPeptMPDB v1.2, respectively.

Figure~\ref{ModifiedPeptide}(g) presents a cross-entropy analysis evaluating the confidence levels of different models in predicting labels, demonstrating that Bi-TEAM exhibits significant advantages across multiple membrane permeability datasets. The consistently lower mean cross-entropy values—particularly pronounced in the Rezai dataset where the value approaches zero—highlight superior probability calibration. Although we observe higher standard deviations for the ProPAMPA, ProCacoPAMPA, and CycPeptMPDB v1.2 datasets, this primarily reflects the model's decisive confidence profile; it generates sharply polarized predictions rather than the ambiguous probabilities ($\sim 0.5$) typical of uncertain models. This distinct confidence pattern, combined with lower overall cross-entropy, confirms the model's ability to capture meaningful permeability-determining patterns. On the Rezai dataset (comprising all-positive samples), our method achieves 100\% prediction accuracy—a 40\% increase over state-of-the-art frameworks—underscoring the model's strong potential in computational membrane-permeability prediction for drug discovery pipelines. By synergistically integrating both protein and chemical language models, Bi-TEAM effectively combines localized chemical insights with a global view of peptide functionality, offering a practical route for designing efficient modified peptides with enhanced membrane permeability.

\subsubsection{Decoding PTMs: Capturing Chemical Shifts in Biological Contexts}

Beyond our extensive experiments on modified peptides formed via non-canonical amino acids, we further validated the generalization capabilities of our proposed model on Post-Translational Modification (PTM) datasets. Here, we focused specifically on the druggability prediction task, which assesses PTM sequences that influence therapeutic targetability, focusing on how modifications alter protein structure and the accessibility of binding sites \cite{li2022dbptm, peng2024ptm, peng2025ptm}.

To illustrate the underlying mechanism, we present phosphorylation as a representative example of PTM function (see the rightmost column of Fig.~\ref{NaturalPTM}(b)). During phosphorylation, a kinase utilizes ATP as a phosphate donor to transfer a phosphate group onto a specific substrate. This process generates ADP and a phosphorylated substrate bearing a covalent phosphate group, thereby altering the substrate's activity or function. In general, various PTMs operate through specific enzymes that transfer distinct chemical groups or small proteins from donor molecules to specific sites on a protein, achieving dynamic regulation of conformation and function.

Fig.~\ref{NaturalPTM}(a) and (b) present a comparative analysis of different methods under random-split and similarity-split scenarios, respectively. Given that the leading PTM-specific algorithm PTMMamba is built on the 650M-parameter ESM-2 model, whereas Bi-TEAM by default uses the 150M-parameter ESM-2 variant, our comparison primarily focuses on configurations based on these two ESM-2 scales.

In the random-split scenario, Fig.~\ref{NaturalPTM}(a) visually demonstrates Bi-TEAM's dominant performance across all metrics. Notably, when utilizing the same 650M-parameter protein encoder, Bi-TEAM\_650M achieves relative improvements of approximately 7\% in MCC and 6\% in Accuracy compared to PTMMamba. The superior Accuracy indicates that the model provides more precise predictions across the global sample set, offering a more comprehensive and reliable distinction between potentially druggable and non-druggable PTM sites. The substantial increase in MCC suggests that Bi-TEAM not only increases the number of correct predictions but also effectively reduces both false positives and false negatives, maintaining a superior decision boundary between positive and negative samples. Furthermore, Bi-TEAM\_650M achieves consistent improvements in ROC-AUC, Precision, and F1-score—even atop already high baselines—pushing the absolute values of all three metrics close to 0.95. A higher ROC-AUC reflects the model's superior ranking capability in distinguishing druggable sites, while the synchronous rise in Precision and F1-score demonstrates excellent sample coverage with minimized false alarms. Collectively, these fine-grained improvements over high baselines further attest to the validity and reliability of Bi-TEAM in characterizing PTM druggability features and enhancing the quality of candidate targets.

Encouragingly, while the scaling law remains applicable within the same algorithm, Bi-TEAM, based on the smaller ESM\_150M consistently outperforms other comparison algorithms based on the larger ESM\_650M in terms of Accuracy and Precision, effectively achieving "more with less." While PTMMamba shows improvement over the standard ESM baseline of the same scale by incorporating PTM-type tokens, the gain is limited. This suggests that merely expanding biological space information via simple token addition yields marginal benefits for this task, whereas incorporating fine-grained chemical information proves to be the superior strategy.

We further conducted comparative experiments using a similarity-split strategy for train-test partitioning. The rightmost t-SNE visualizations in Fig.~\ref{NaturalPTM}(b) clearly illustrate the distinct distribution shift between the selected test set (red cross) and the training set (black crosses). Under these more rigorous and realistic conditions, the proposed method demonstrates an overwhelming advantage across the board, as visually evidenced by the radar charts, bar graphs, and line plots. Specifically, relative to the best baseline, Bi-TEAM\_650M achieved relative improvements of over 18\% in MCC and 9\% in Accuracy, further widening the performance gap. The primary driver of this success is Bi-TEAM's ability to maintain high performance under strict data partitioning, whereas other comparison algorithms suffered widespread performance degradation. This resilience is mirrored in the other three metrics, serving as strong proof of the proposed algorithm's high generalization capability. Additionally, while the scaling law persists, the nearly identical performance of PTMMamba and ESM\_650M indicates that out-of-distribution testing further diminishes the utility of PTM-type tokens. To further demonstrate the generalizability of our algorithm to other PTM-related downstream tasks, we provide predictions of PTM–disease associations in Appendix~\ref{SI_PTM}.

\subsubsection{Validating on Natural Peptides and Proteins: Robustness via Cross-Modal Learning}

Having validated the efficacy of Bi-TEAM on modified peptide and PTM datasets, we further assessed its cross-domain generalization capabilities on natural protein datasets, specifically focusing on solubility and hemolysis prediction tasks. To provide a mechanistic basis for these predictions, Fig. \ref{NaturalPTM}(h) and (i) illustrate the key steps and interaction dynamics governing the peptide hemolytic process and protein solubility changes, offering an intuitive understanding of the biological underpinnings for these properties. Fig. \ref{NaturalPTM}(c-d) presents a comparative analysis of Bi-TEAM against various baseline models under a random train-test split strategy, while Fig. \ref{NaturalPTM}(e-f) details performance under a more rigorous similarity-based split strategy. Generally, compared to modified peptide scenarios that lack large-scale training data and well-established evolutionary patterns, ESM—pre-trained on vast natural protein sequence databases—typically exhibits superior sequence–function predictive capabilities. Consequently, ESM's performance on standard amino acid-based datasets surpasses that of chemical language models across most metrics. 

Remarkably, despite the dominance of evolutionary baselines in this domain, Bi-TEAM consistently achieves state-of-the-art performance. This superiority indicates that chemical representations provide orthogonal value even for natural peptides and proteins, capturing explicit physicochemical drivers that complement the evolutionary context. Unlike naive feature concatenation, which risks introducing redundancy, Bi-TEAM dynamically modulates the injection of fine-grained chemical insights into the broad biological semantic space. By integrating chemical features as a precise refinement mechanism, the model effectively bridges the gap between macro-evolutionary patterns and micro-level physicochemical determinants.

For the hemolysis prediction task, where the dataset exhibits a severe class imbalance (positive-to-negative ratio of 1:4) compared to ProPAMPA, the MCC metric serves as a critical performance indicator. Under random splitting, Bi-TEAM achieves an MCC score 2.5 times greater than that of ESM, indicating a superior capability to reduce both false negatives and false positives for balanced classification. Moreover, Bi-TEAM exceeds ESM by over threefold in F1-score, revealing a significantly improved balance between true positive recognition and precision. An improvement of 12.8\% in ROC-AUC, alongside noticeable gains in Accuracy, further underscores Bi-TEAM’s enhanced discriminative ability across different thresholds. While the increases in Accuracy and Precision are relatively smaller, they reach state-of-the-art levels, demonstrating comprehensive improvements across all confusion matrix elements. In contrast to PLMs, CLMs operate within a chemical space constructed from chemical tokens. However, as shown in Fig. \ref{ModifiedPeptide}(c) and Fig. \ref{NaturalPTM}(c), the performance of SELF-BART and ChemBERTa diverges substantially. ChemBERTa’s MCC, Precision, and F1-score collapse to zero, suggesting a complete failure to distinguish between hemolytic and non-hemolytic peptides. While SELF-BART improves upon this by adopting SELFIES-based embeddings to avoid structural representation errors, it still lags behind other deep learning methods, likely due to domain gaps inherent in applying small-molecule pre-trained models to peptide tasks. These findings suggest that limiting exploration to a single-scale space often yields unsatisfactory results, whereas cross-scale biochemical space representations emerge as the more effective approach. Compared to Random Forest (the second-best model), Bi-TEAM demonstrates a 17.4\% relative improvement in F1-score and an 18.8\% improvement in MCC, highlighting a stronger balance in per-class predictions.

Under the more stringent similarity-based train-test partitioning strategy (Fig. \ref{NaturalPTM}(e)), the proposed algorithm maintains a commanding lead. Bi-TEAM achieves an astounding 350\% improvement in MCC over the best baseline, SELF-BART, whereas the MCC for ESM and ChemBERTa degrades to zero. In terms of ROC-AUC, Precision, and F1-score, Bi-TEAM demonstrates noteworthy relative improvements of 16.33\%, 42.60\%, and 252\%, respectively. Although its Accuracy is marginally lower than ESM and ChemBERTa, those baselines fail to record valid values for MCC, Precision, and F1-score, while SELF-BART achieves only a fraction of Bi-TEAM's performance. We hypothesize that the severe imbalance (1:4) in the Hemolysis dataset—compared to the milder 2:1 ratio in ProPAMPA—causes baselines to prioritize majority class accuracy at the expense of discriminative power. Under these rigorous splits, existing models struggle to predict properties for proteins with diverse structural characteristics, whereas Bi-TEAM manifests striking robustness and cross-structural generalization capacity.

Regarding the solubility prediction task, Fig. \ref{NaturalPTM}(d) presents comparative results under random splitting. Bi-TEAM significantly leads in key indicators, achieving improvements of 8.4\% in MCC, 3.0\% in F1-score, and 2.3\% in Precision compared to the best baseline, ESM. Traditional machine learning models (DT, KNN, RF, SVM) generally fail to approach the performance of Bi-TEAM or ESM in Accuracy or ROC-AUC and exhibit greater sensitivity to data partitioning. Fig. \ref{NaturalPTM}(f) further illustrates performance under the similarity-split strategy (see the rightmost column for train-test split visualization), where the radar chart highlights the comprehensive advantage of the proposed algorithm. Bi-TEAM shows consistent improvements of 15.4\%, 4.5\%, 3.8\%, 4.1\%, and 4.3\% over the best baseline in MCC, Accuracy, ROC-AUC, Precision, and F1-score, respectively. Here, ESM outperforms ChemBERTa and SELF-BART, attributed to its large-scale pre-training that effectively captures long-range interactions and the distinct clustering of the Solubility dataset, which allows ESM's strengths to be fully harnessed. Even so, Bi-TEAM retains its inherent strengths in capturing global sequence semantics and fine-grained chemical details, delivering the best performance. Additional comparisons for the protein non-fouling prediction task are provided in Appendix \ref{natural}.

Overall, Bi-TEAM not only excels under random data partitioning but also maintains outstanding performance under stringent similarity-aware splits, underscoring its ability to handle molecular structural diversity and class imbalance while demonstrating robust extrapolation capabilities across hemolysis, solubility, and non-fouling properties.

\subsection{Unlocking Uncharted Chemical Space Guided by Bi-TEAM}

\begin{figure}[htbp!]
    \centering
\includegraphics[width=1.0\textwidth]{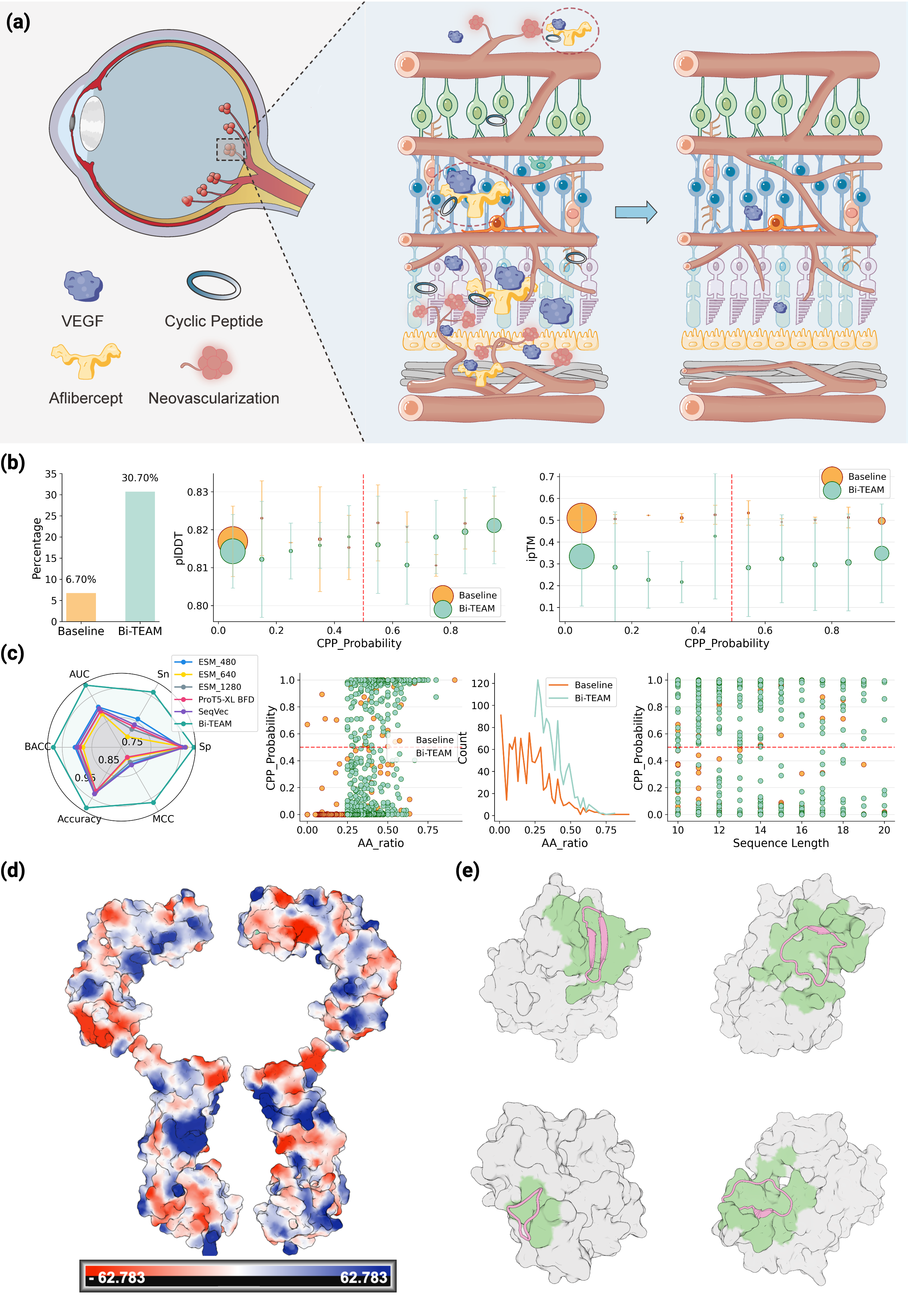} 
    \caption{\textbf{Design of a non-invasive drug delivery system.} \textbf{(a)} A schematic diagram of the ocular absorption pathway of the AFL/cyclic peptide complex: The red dashed circle represents the binding of the cyclic peptide-aflibercept complex to VEGF. After reaching the fundus, aflibercept is released to treat neovascularization, and the cyclic peptide will be degraded in serum or ocular tissue. \textbf{(b)} The success rate of the generated 1000 samples and the corresponding pLDDT and ipTM distributions. \textbf{(c)} From left to right, the performance radar chart of the cell-penetrating classification benchmark, the relationship between key hydrophobic amino acids and cell penetration probability and sample count for the 1000 generated samples, and the relationship between cyclic peptide length and cell penetration probability. \textbf{(d)} Aflibercept was displayed in electrostatic surface potentials colored red (-) and blue (+). \textbf{(e)} The structure of four top-ranked AFL/cyclic peptide complexes.}
    \label{Gen}
\end{figure}
Current state-of-the-art generative models, whether diffusion-based approaches like RFdiffusion\cite{watson2023novo} or hallucination-based methods such as BindCraft\cite{pacesa2024bindcraft} and BoltzDesign1\cite{cho2025boltzdesign1}, have demonstrated remarkable success in designing protein binders with high binding affinity. However, these models primarily optimize structural metrics through their training losses or gradient objectives, often overlooking critical constraints related to specific physicochemical properties. Consequently, the generated binders, while structurally sound, may lack the necessary physicochemical characteristics for real-world biological applications. To address this limitation, we leverage Bi-TEAM, which performs joint learning across both biological and chemical spaces to effectively represent complex physicochemical property landscapes, to navigate the generative process. Therefore, we can guide the design of binders with tailored properties, which is a capability that remains beyond the reach of conventional generative models.

The following sections will delve into a specific generative case study: the design of a non-invasive drug delivery system. We will first introduce the background, then train and evaluate the Bi-TEAM model to predict cell-penetrating non-canonical cyclic peptides, and finally, based on the hallucination method, utilize Bi-TEAM as an additional gradient guide to lead the generation and analyze the generation results.

\subsubsection{Background and Motivation}
Neovascular age-related macular degeneration (nAMD) represents a primary cause of irreversible blindness among the elderly population. The core pathology involves choroidal neovascularization and vascular leakage driven by pathological vascular endothelial growth factor (VEGF)\cite{apte2021age,fleckenstein2024age}. Currently, the first-line clinical therapy relies on intravitreal injections of large-molecule anti-VEGF agents, such as Aflibercept, to neutralize VEGF and inhibit angiogenesis\cite{wecker2017five}. However, Aflibercept possesses a substantial molecular weight (115 kD), rendering it unable to penetrate complex ocular physiological barriers—including the cornea, conjunctiva, and sclera—to reach the retina via simple topical administration. Furthermore, long-term, frequent intravitreal injections impose a significant burden on both patients and healthcare systems and are associated with risks of severe complications, such as endophthalmitis and retinal detachment. These factors contribute to poor patient compliance, leading to treatment interruption and subsequent vision loss\cite{ciulla2022longer}. Therefore, designing a peptide binder capable of specifically binding Aflibercept and facilitating its transport across ocular barriers is critical for the development of non-invasive eye drop therapies. A schematic illustration of this therapeutic strategy and the challenges it addresses is presented in Figure \ref{Gen}a.

Nevertheless, traditional simple linear peptides face insurmountable physiological and physicochemical bottlenecks in ocular delivery applications\cite{fan2025peptide}. First, the ocular surface tear fluid is rich in various proteases and peptidases. Due to their exposed N and C-termini, linear peptides are rapidly recognized and hydrolyzed, resulting in an extremely short half-life on the ocular surface and an inability to maintain effective concentrations for cross-barrier transport. Second, linear peptides typically exhibit disordered, "flexible" conformations in solution, possessing high conformational entropy. Binding to macromolecular drugs like Aflibercept requires overcoming a massive entropy loss to fix the conformation; this often results in weak binding affinity and insufficient specificity, making it difficult to form stable "carrier-drug" complexes.

In contrast, cyclic peptides—formed by chemically bonding the peptide chain ends or side chains—demonstrate significant advantages. Structural rigidity is a core characteristic of cyclic peptides; this pre-organized conformation reduces the entropy penalty upon binding, enabling the peptide to engage Aflibercept with superior affinity and selectivity. Furthermore, by reducing the exposed polar surface area and facilitating the formation of intramolecular hydrogen bonds, cyclic peptides significantly enhance transmembrane permeability. Crucially, the cyclized structure eliminates free termini and utilizes steric hindrance to effectively resist ocular enzymatic degradation, thereby drastically improving metabolic stability. Consequently, designing a cyclic peptide binder with specific conformational constraints not only resolves the issues of poor stability and weak binding associated with linear peptides but also represents a key breakthrough for achieving efficient, non-invasive posterior segment delivery of Aflibercept.

\subsubsection{Cell-Penetrating Peptide Prediction and Evaluation}
The construction of the dataset follows the standard protocol established by pLM4CPP\cite{kumar2025plm4cpps}. The data is primarily integrated from authoritative databases including CPPsite2.0, C2Pred, CellPPD, MLCPP 2.0, and KELM-CPPpred. In this dataset, positive samples are defined as experimentally validated cell-penetrating peptides (CPPs), while negative samples consist of sequences confirmed by experiments or reported in the literature to lack penetrating activity. After rigorous screening and redundancy removal, the final dataset contains 1,399 positive samples and 4,080 negative samples.

To evaluate the predictive performance of Bi-TEAM for cell-penetrating peptides, we conducted a comprehensive comparison against state-of-the-art embedding models, including SeqVec\cite{heinzinger2019modeling}, ESM2\cite{lin2023evolutionary}, and ProtT5 variants\cite{elnaggar2021prottrans}. Bi-TEAM achieved state-of-the-art performance across all six evaluation metrics, significantly outperforming the second-best models in each category which is shown in Figure \ref{Gen}c: it surpassed the runner-up in Accuracy (ACC) by 5.52\% (0.9872 vs. 0.932 from SeqVec), Balanced Accuracy (BACC) by 5.88\% (0.9822 vs. 0.907 from ESM2-480), Sensitivity (Sn) by 12.58\% (0.9658 vs. 0.860 from ESM2-480), Specificity (Sp) by 1.45\% (0.9915 vs. 0.977 from ProtT5-XL BFD), MCC by 14.68\% (0.9658 vs. 0.819 from SeqVec), and AUC by 8.45\% (0.9915 vs. 0.907 from ESM2-480). The substantial improvement in sensitivity (+12.58\%) and MCC (+14.68\%) is particularly noteworthy, indicating that Bi-TEAM is exceptionally robust at correctly identifying true positive CPPs while maintaining a superior balance between precision and recall compared to existing methods.

\subsubsection{Attribute-Guided Generation and Evaluation}
We employed BoltzDesign1 as a baseline framework to design cyclic peptides targeting Aflibercept. Specifically, we generated 1,000 cyclic peptides with lengths ranging from 10 to 20 amino acids under two conditions: BoltzDesign1 with default structural constraints and BoltzDesign1 augmented with Bi-TEAM guidance. The success rate was subsequently evaluated using Bi-TEAM, with success defined as a prediction logit greater than 0.5. To demonstrate the directed shift in chemical space sampling under Bi-TEAM guidance, We analyzed the amino acid sampling preferences in samples generated by Bi-TEAM and examined whether these preferences align with known amino acid biases that favor membrane permeation. 

As illustrated in Figure \ref{Gen}b, traditional structure-based generation methods struggle to produce cell-penetrating cyclic peptides, yielding a success rate of only 6.7\%. This suggests that cell-penetrating cyclic peptides likely occupy rare regions within the sampling distribution of standard generative models. However, Bi-TEAM guidance enables the structural generative model to explore a significantly broader chemical space, substantially increasing the success rate to 30.7\%. Crucially, this dramatic improvement was achieved without compromising default structural constraints, including overall structural confidence (pLDDT), binding interface confidence (ipTM), and helical secondary structure requirements. As shown in the middle panel of Figure \ref{Gen}b, even in sampling regions with high CPP probabilities, the generated peptide-Aflibercept complexes maintained an average pLDDT score exceeding 0.82.

We conducted an in-depth analysis to elucidate the mechanisms underlying the superior success rate of cyclic peptide binders generated via Bi‑TEAM–guided BoltzDesign1. Previous studies on cell-penetrating peptides have established that a classic hydrophobic triad—Tryptophan (W), Phenylalanine (F), and Tyrosine (Y)—plays a critical role in membrane translocation \cite{fan2025peptide}. In addition, positively charged amino acids, particularly Arginine (R) and Lysine (K), are known to further facilitate membrane interaction and penetration.

By analyzing the occurrence patterns of these key residues across 1,000 generated samples, we observed that Bi‑TEAM guidance induces a systematic shift in the chemical space sampled by BoltzDesign1, steering peptide generation toward sequences that are more compatible with experimentally validated cell-penetrating motifs. Specifically, as shown in the middle panel of Figure \ref{Gen}c, peptides generated under Bi‑TEAM guidance exhibit a significantly higher co-occurrence frequency of the hydrophobic triad (W/F/Y) together with two positively charged residues (R/K) compared to the original, unguided generation. We observed a similar trend when analyzing the distribution of residue counts. This enriched distribution directly aligns with known structure–function principles for cell-penetrating peptides, indicating that Bi‑TEAM not only captures the qualitative mechanism of the hydrophobic triad but also quantitatively enhances the likelihood of generating sequences with membrane-permeable characteristics. Finally, we ruled out the influence of peptide length (10–20 residues) as a confounding factor: we found no significant correlation between peptide length and CPP probability. Collectively, rather than uniformly exploring the unconstrained sequence space, Bi‑TEAM effectively biases the sampling distribution toward an appropriate “target region” in the chemical–biological joint space, where hydrophobic and electrostatic features are optimally balanced for membrane translocation.

\subsubsection{Case Studies}
We first visualize the three-dimensional structure of the Aflibercept dimer in Figure \ref{Gen}d, where the molecular surface is colored according to electrostatic potential (red for negative, blue for positive). Subsequently, we employed AlphaFold3 to predict the complex structure of the designed cyclic peptide bound to aflibercept, allowing for the identification of specific interfacial residues (Figure \ref{Gen}e).

Analysis of the Aflibercept R1D2 domain surface revealed two distinct binding pockets accommodating the generated cyclic binder. The first pocket is a hydrophobic cavity defined by three loops: Loop 1 (Leu46–Leu49), Loop 2 (Asn91–Arg96), and Loop 3 (Val132–Asp135). The second binding pocket is formed by a loop region spanning residues Asp52–Asp59 and a $\beta$-sheet segment comprising residues Phe64–Ser67.

\subsection{Ablation Studies}

\subsubsection{Performance comparison between full model and variants}

\begin{figure}[htbp]
    \centering
        \includegraphics[width=0.9\textwidth,height=0.9\textheight]{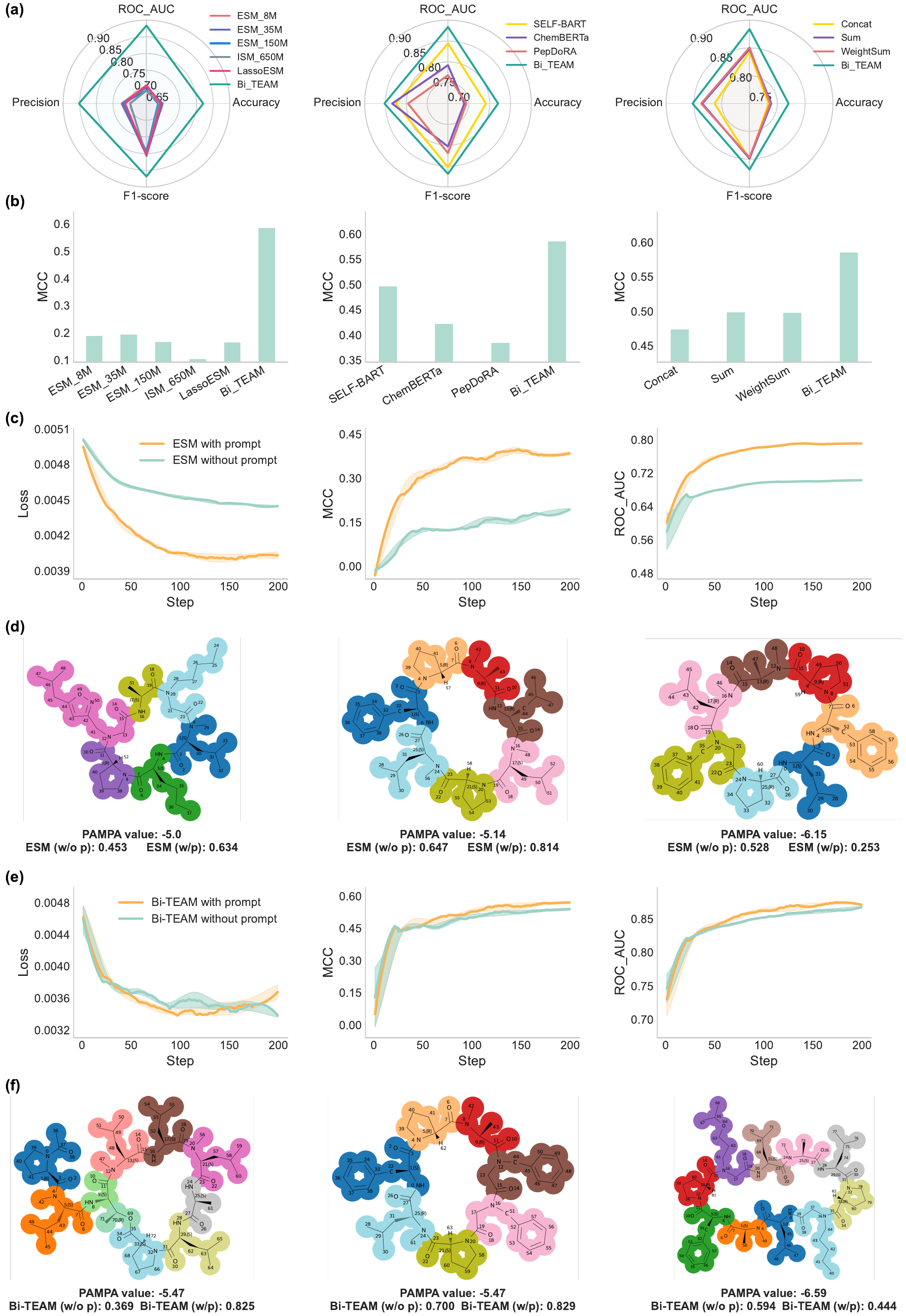}
    \caption{\textbf{Ablation study results.}
    \textbf{(a)(b)} Compare four metrics and MCC, respectively, for different PLMs, CLMs, and fusion methods (from left to right);
    \textbf{(c)(e)} Present ESM-based and Bi-TEAM models (with/without prompt) in terms of training loss, MCC, and ROC\_AUC (from left to right);
    \textbf{(d)(f)} Shows three sets of example predictions for both ESM-based and Bi-TEAM-based models (with/without prompt).}
    \label{ablationStudy}
\end{figure}
Bi-TEAM integrates both PLM and CLM for a more comprehensive exploration of peptide space. To investigate the impact of different components on the overall model, we provide comparative experiments of various PLMs, CLMs, and fusion strategies.

As illustrated in Fig.~\ref{ablationStudy}(a), for PLMs (first column), although ESM\_35M outperforms ESM\_8M, further scaling to ESM\_150M or ISM\_650M does not improve generalization for non-canonical peptides and may even increase the risk of overfitting. Compared to the strongest PLM baseline, ESM\_35M, Bi\_TEAM achieves relative improvements of approximately 26.4\%, 17.4\%, 18.1\%, and 7.7\% in ROC\_AUC, accuracy, precision, and F1-score, respectively. For CLMs (second column), while SELF-BART is the leading single-modality model, Bi\_TEAM still surpasses it with relative improvements of approximately 4.7\%, 3.5\%, 3.3\%, and 1.7\% in ROC\_AUC, accuracy, precision, and F1-score. Although LassoESM utilizes lasso peptide data to fine-tune ESM and possesses coarse-grained topology-aware capabilities, these features do not strongly correlate with membrane permeability, resulting in no significant performance improvements. PepDora, fine-tuned based on ChemBERTa, defines an inappropriate regression task and lacks sufficient training data, causing the fine-tuned model to shift toward a suboptimal representation space and consequently perform worse than ChemBERTa when transferred to membrane permeability classification. Regarding fusion strategies (third column), Sum slightly outperforms Concat and WeightSum; however, Bi\_TEAM still yields notably better performance across different metrics compared to this best fusion baseline, confirming that deeper cross-domain interactions deliver superior results across all key indicators.

Furthermore, Fig.~\ref{ablationStudy}(b) shows that MCC also attest to Bi\_TEAM's advantages. By establishing complementary information flows between PLMs and CLMs, Bi\_TEAM retains crucial evolutionary context from native peptides while capturing fine-grained atomic and functional-group–level chemical features. Against the best PLM baseline, ESM\_35M, Bi\_TEAM achieves nearly a twofold relative improvement in MCC and approximately 17\% compared to SELF-BART. In contrast, simpler fusion methods such as Concat, Sum, or WeightSum yield only moderate MCC gains. Specifically, Bi\_TEAM demonstrates approximately a 20\% relative improvement in MCC over the best fusion baseline. Notably, SELF-BART, built upon the SELFIES format, encodes cyclic and branched structures more robustly, mitigating ambiguity from topological variances. Integrating PLM evolutionary signals with CLM-based chemical representations thus enables a more thorough characterization of ncAAs and confers performance advantages across multiple challenging metrics.

\subsubsection{Validation of prompt effectiveness}
We introduce the Modification Location Prompt into non-canonical peptide modeling with the core motivation of providing explicit positional information that marks which residues in the sequence are non-canonical. By inserting a dedicated location token that signals exactly ``where" the ncAA has been introduced, we offer an additional level of focus for the model on residues whose functional or structural role is critical. This context-aware prompt ensures the surrounding chemical tokens are given higher attention whenever local modifications may significantly influence broader peptide properties. Inspired by using estimated antigen-binding sites as prompts for Nanobody--antigen interaction prediction\cite{deng2024nanobody}, we regard the explicit identification of ``ncAA position" as a form of high-level chemical information that helps overcome the limitation of original ESM approaches, which primarily rely on substituting ncAA with the most similar natural amino acids for representation.

To validate the effectiveness of the Modification Location Prompt, we performed ablation studies on the ProPAMPA dataset and present comparative results for the ESM and our proposed method with and without prompts in Fig.~\ref{ablationStudy}(c)(d) and Fig.~\ref{ablationStudy}(e)(f). From the training loss, MCC, and ROC\_AUC curves in Fig.~\ref{ablationStudy}(c), we observe that the usage of the prompt not only accelerates convergence for the ESM but also significantly outperforms the non-prompt counterpart on key metrics. Without a prompt, the original ESM can only approximate ncAAs using the most comparable natural amino acids, leading to suboptimal predictions for ncAA-containing sequences. In contrast, explicitly highlighting ncAA positions enables ESM to better capture the influence of the modified residue on overall peptide behavior. Although Bi-TEAM already achieves superior performance by integrating chemical- and bio-spatial information of ncAAs, Fig.~\ref{ablationStudy}(e) shows that incorporating the Modification Location Prompt still provides a consistent improvement across different metrics. This underscores the fact that signaling the ncAA position faithfully reflects local chemical modifications.

To more intuitively demonstrate the effectiveness of the Modification Location Prompt, we present three sets of examples for both ESM and Bi-TEAM in Fig.~\ref{ablationStudy}(d)(f). In Fig.~\ref{ablationStudy}(d), the left subplot shows a permeable non-canonical peptide (PAMPA value $>$ -6), yet the ESM model without a prompt incorrectly predicts it as non-permeable (prediction probability: 0.453). By contrast, incorporating the prompt raises the prediction probability to 0.634, correctly matching the ground truth (predictions with probability $>$ 0.5 are labeled as positive). In the middle subplot, even when both variants make a correct call, the prompt further increases the model's confidence; whereas on the right, the true label is negative, and the ESM model without a prompt incorrectly labels it as positive while the prompted version correctly assigns a negative prediction. A similar set of results using Bi-TEAM is shown in Fig.~\ref{ablationStudy}(f). In Appendix \ref{appendix:caseStudy}, we provide additional examples comparing predictions from both ESM and Bi-TEAM without and with prompts. 

Through this comprehensive prompt ablation study—combining multiple metrics with concrete examples—we demonstrate the substantial impact and necessity of a carefully designed prompt in driving overall model performance. Furthermore, we have conducted an in-depth exploration of the contribution of each module within our proposed Bi-TEAM architecture, and comprehensive experimental findings are provided in Appendix \ref{appendix:ablationStudy}.

\subsubsection{Space Modification Analysis from the Perspective of Rank}

To investigate how our fusion method combines information from protein language models and chemical language models, we conducted a rank analysis using Singular Value Decomposition (SVD). We introduced two key metrics: the Max-over Ratio (MoR), which measures the ratio of the largest single-modality rank to the fused rank, and the Sum-coverage Ratio (ScR), which compares the fused rank to the sum of individual modality ranks. By applying different thresholds to filter singular values, we estimated the effective ranks of PLM embeddings, CLM embeddings, and their fused representation.

Our experiments (detailed in Appendix~\ref{appendix:rank}) reveal distinct patterns across different scenarios. In scenarios without modification location prompts, both MoR and ScR remain below 1, with the fused rank consistently falling between the maximum individual rank and the sum of both ranks. This pattern reflects three fundamental principles of multimodal fusion: first, when modalities contain completely different information, the fused rank approaches their sum; second, when modalities share overlapping features, redundancy prevents simple addition of ranks; third, and most importantly, the fusion process creates cross-modal interactions that elevate the rank beyond any single modality alone, demonstrating clear benefits of multimodal integration.

When modification location prompts are introduced, two notable changes emerge: ScR can exceed 1 at higher thresholds, while MoR drops substantially. These shifts indicate that prompts introduce additional information that expands the representation space beyond what PLM and CLM originally capture. Across both scenarios, higher thresholds consistently increase ScR while decreasing MoR, suggesting that stricter filtering removes noise and retains only the most meaningful singular values for more reliable rank estimation. This complementary behavior between ScR and MoR further confirms that our fusion method produces high-quality representations by effectively integrating complementary information from multiple modalities. For the extended methodology and additional results for rank analysis, see Appendix~\ref{appendix:rank}.

\subsection{Feature Interpretation}

\begin{figure}[h!]
    \centering
     \includegraphics[width=\textwidth]{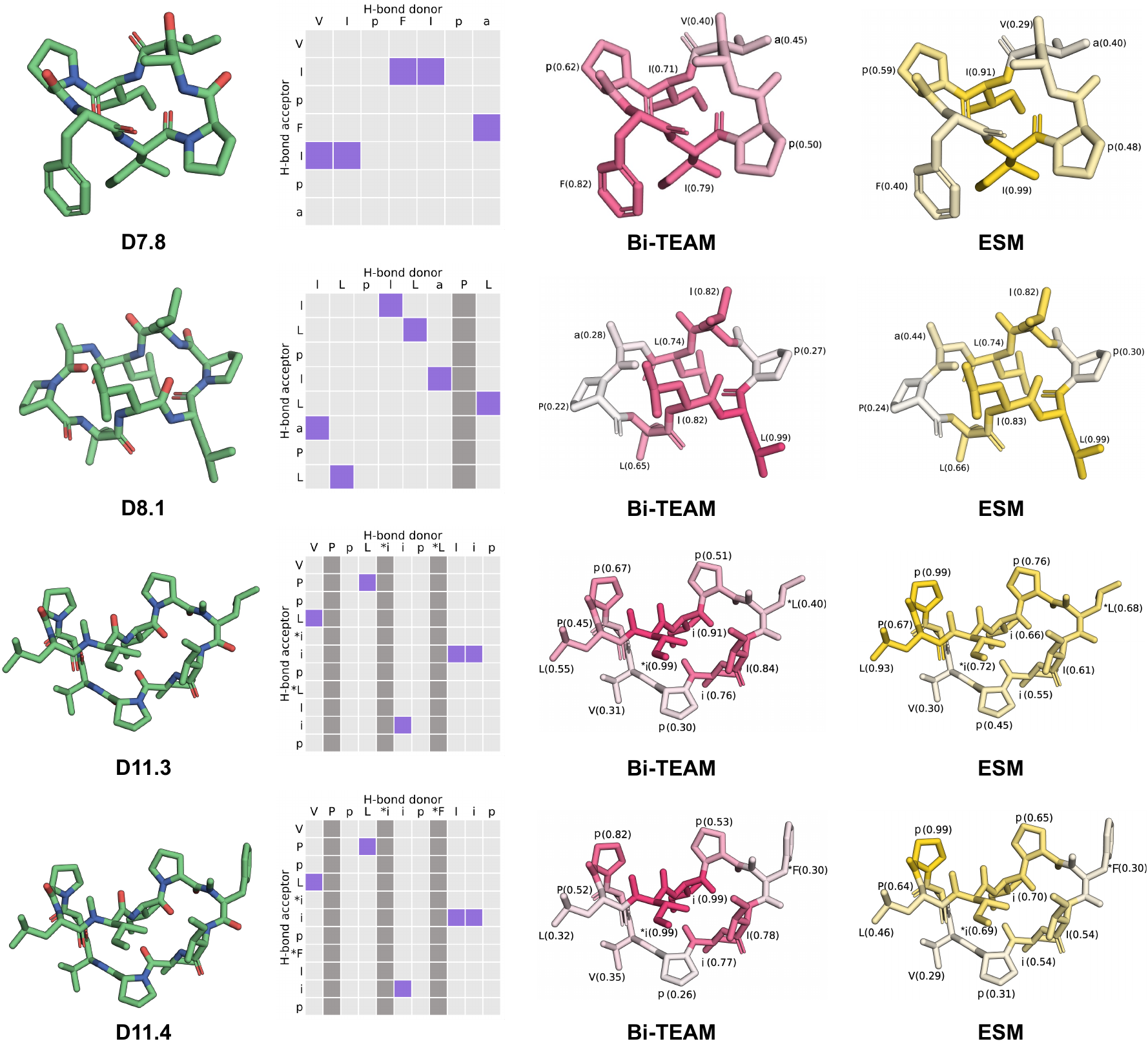}
    \caption{\textbf{Feature interpretation.} From left to right: structure of four designed cyclic peptides, hydrogen bond donor-acceptor analysis, attention weight visualization from the Bi-TEAM model, and attention weight visualization from the ESM model.}
    \label{h-bond}
\end{figure}

Deep learning models have demonstrated remarkable generalization capabilities in various biological and chemical tasks, prompting ever-growing interest in understanding their internal decision-making mechanisms. A deeper examination of these factors not only helps bridge conventional chemical knowledge and the patterns learned by deep models, but also reveals novel interpretative insights and design strategies that can inform targeted drug discovery.

Previous studies have shown that intermediate representations in language models often reflect patterns that correspond to established knowledge in structural biology or quantum chemistry \cite{simon2024interplm}. For example, PLMs have been used to detect secondary structures or protein binding pockets \cite{zhang2024protein,simon2024language}, and attention maps have revealed protein contact relationships \cite{vig2020bertology}. Further combining techniques like integrated gradients enables identification of active sites and transmembrane regions at various scales \cite{marks2011protein}. However, these existing methods focus primarily on a single modality or a single scale of representation, leaving the underlying mechanisms of how multi-modal information is fused and how different modules leverage this multi-modal information to accomplish tasks like transmembrane permeation prediction largely unexplored.

To validate whether the model indeed captures traditionally recognized molecular modifications such as N-methylation, intramolecular hydrogen bonding, and cis/trans conformations and to quantitatively assess their relative impact on transmembrane permeability for improved molecular design and subsequent drug optimization \cite{buckton2021cyclic}, we adopt a saliency map-based method to quantify the importance of each amino acid token \cite{simonyan2013deep}. Specifically, for each sample in the test set, we compute the gradient of the cross entropy loss with respect to the amino acid token embeddings, without updating the model parameters. These gradients serve solely to gauge token-level importance. We then obtain the ESM model embeddings from the final three layers and apply mean pooling across layers. The final importance score is calculated by multiplying the pooled embedding with the corresponding token gradient:
\begin{align}
\bar{\mathbf{e}}_{i} &= \frac{1}{3}\sum_{l=L-2}^{L} \mathbf{e}_{i}^{(l)} \\
\mathbf{s}_{i} &= \nabla_{\mathbf{x}_{i}} \mathcal{L} \cdot \bar{\mathbf{e}}_{i}
\end{align}
where $\bar{\mathbf{e}}_{i}$ is the average embedding representation of the $i$-th token from the last three layers ($L, L-1, L-2$), $\mathbf{e}_{i}^{(l)}$ is the embedding representation of the $i$-th token at layer $l$, $L$ is the total number of layers in the model. $\nabla_{\mathbf{x}_{i}} \mathcal{L}$ is the gradient of the loss function $\mathcal{L}$ with respect to the input embedding of the $i$-th amino acid token, $\mathbf{s}_{i}$ is the importance score for the $i$-th amino acid token.

To demonstrate the generalizable knowledge captured by our model, we evaluated its performance on an external test set validated by wet-lab experiment, ProCacoPAMPA \cite{bhardwaj2022accurate}. This dataset comprises de novo designed transmembrane cyclic peptides containing non-canonical amino acids. Researchers successfully crystallized and solved high-resolution X-ray crystal structures for 15 macrocycles, of which 12 structures exhibited close structural alignment with the designed conformations. Fig.~\ref{h-bond} presents a comparative analysis with the leftmost column showing the X-ray crystallographic structures, the second column illustrating hydrogen bond donors and acceptors within the peptide structure, the third column depicting importance scores assigned by our Bi-TEAM to different amino acid residues, and the rightmost column showing importance scores from ESM. 

Following the analytical framework established in \cite{bhardwaj2022accurate}, we focused on three critical structural determinants: N-methylation, intramolecular hydrogen bonding networks, and cis-trans isomerization states.

N-methylation of the peptide backbone represents a critical chemical modification associated with enhanced pharmacological properties of cyclic peptides. This modification alters the conformational landscape, hydrogen bonding potential, and lipophilicity of the macrocycle, thereby facilitating membrane permeability. Introduction of a methyl group to the peptide backbone increases molecular hydrophobicity and influences the cis-trans equilibrium of N-methylated amide bonds, thus enhancing the molecule's chameleonic capacity. 

Intramolecular hydrogen bonding also emerges as a critical determinant of macrocycle membrane permeability. These sophisticated intramolecular interactions enable complete sequestration of otherwise exposed NH donors, effectively removing polar functionalities from solvent exposure without requiring extensive N-methylation. High-resolution crystallographic studies of the permeable macrocycles consistently demonstrate elaborate hydrogen bonding architectures that effectively mask polar backbone functionalities that would otherwise create insurmountable energetic barriers to membrane traversal. The presence of these internal hydrogen bond networks constitutes the predominant structural feature governing membrane permeability and represents a fundamental design principle for engineering membrane-permeable peptides.

Furthermore, while cis-peptide bonds alone do not substantially enhance permeability, cis-trans isomerization facilitates membrane permeability by enabling conformational interconversion between distinct states. In nonpolar milieus, some peptides preferentially adopt conformations with fewer or no exposed unsatisfied NH donors, while in polar environments, conformations with greater NH exposure predominate. This chameleonic behavior enables peptides to expose backbone NH groups in one state—potentially facilitating target engagement—while maintaining significant permeability through conversion to a hydrogen bond-saturated state during membrane traversal. 

Then we divide the experimental results presented in Fig.~\ref{h-bond}  into two groups for analysis: peptides 7.8 and 8.1 primarily investigate the effects of intramolecular hydrogen bonding and cis/trans conformations, while peptides 11.3 and 11.4 focus on the impact of N-methylation. We mark natural amino acids as uppercase, amino acids with cis-peptide bonds as lowercase and N-methylated amino acids are preceded by an asterisk (*). To distinguish different amino acids in a sequence, we define the index as the sequence from left to right, starting from 1, and represented by amino acid plus index. For example, I4 represents the fourth 'I' amino acid from left to right.

In D7.8, the residues I2, F4, and I5 attract the highest attention from our model, with importance scores of 0.71, 0.82, and 0.79, respectively. These three amino acids are the principal participants in forming intramolecular hydrogen bonds, which presumably contribute to membrane permeability. By contrast, ESM assigns a lower importance score (0.40) to F4, potentially indicating that ESM did not fully capture the critical hydrogen bond (F–a) responsible for membrane permeability. This discrepancy reflects ESM's lower sensitivity to the subtle hydrogen-bonding interactions that can exist between donor–acceptor pairs.

In D8.1, most residues participate in intramolecular hydrogen bonds except p3 and P7. Consistent with this observation, Bi-TEAM assigns minimal importance to these two residues, with scores of 0.27 and 0.22, respectively. Notably, this sequence includes two types of ``p" amino acids (p and P), each capable of cis/trans isomerization, but both fall within low-importance regions (0.27 and 0.22). This observation aligns with previous findings \cite{buckton2021cyclic}, suggesting that the mere existence of cis peptide bonds has negligible correlation with membrane permeability. Interestingly, in this case, ESM captures a pattern similar to that of Bi-TEAM, demonstrating that residue-level information can indeed yield preliminary insights into membrane permeability.

In D11.3, two residues (i5 and L8) are N-methylated. According to our model, the most critical positions are i5, i6, I9, and i10, with importance scores of 0.99, 0.91, 0.84, and 0.76, respectively. All except i5 form intramolecular hydrogen bonds, whereas i5—owing to its N-methylation—achieves a markedly high score without engaging in such bonding. By contrast, L8 (also N-methylated) receives a relatively lower score of 0.40 from Bi-TEAM, indicating that not all N-methylation events equivalently enhance membrane permeability. This pattern corroborates observations reported by \cite{bhardwaj2022accurate}, who noted that certain N-methylations contribute more substantially to permeability than others. However, ESM erroneously attributes an extremely high importance score (0.99) to p3, even though p3 lacks notable features that favor membrane permeability. Moreover, ESM fails to highlight the two functionally important residues—i6 (key hydrogen-bond participant) and i5 (N-methylated)—assigning them only moderate importance scores of 0.70 and 0.69, respectively. This discrepancy suggests that ESM may overlook some subtle but critical interactions that are better captured by Bi-TEAM.

D11.4 is highly similar to D11.3, with the sole difference being the substitution of L8 for F8. Overall, Bi-TEAM maintains a similar importance score distribution, indicating consistent recognition of the structural features that influence permeability. However, it registers a slight decrease in attention to F8 (from 0.40 to 0.30) relative to L8, and also notes a reduction in the subsequent residue I9 (from 0.84 to 0.78). This localized impact demonstrates Bi-TEAM's sensitivity to minor structural modifications. In contrast, the importance scores of ESM were less consistent and could not accurately identify residues i5, i6, I9, and i10 that contributed significantly to membrane permeability.

\section{Discussion}\label{discussion}
Bi-TEAM offers a robust solution for integrating chemical and biological modeling paradigms to more holistically predict the properties of peptides containing non-canonical amino acids (ncAAs) and effectively steer modified peptide generation toward target properties. By establishing the biological space—constructed by Protein Language Models (PLMs)—as the foundational framework for macro-level sequence semantics and then strategically incorporating chemical space features via a Chemical Language Model (CLM), Bi-TEAM selectively activates fine-grained chemical representations precisely when necessary.

This synergistic approach ensures Bi-TEAM operates effectively at multiple scales. It demonstrates that property prediction domains—ranging from modified peptides and Post-Translational Modifications (PTMs) to natural proteins—depend on a delicate interplay between global biological sequence context and local chemical granularity, all without succumbing to the pitfall of token proliferation often seen in purely chemical vocabularies. Its capacity to balance broad biological coverage via PLMs while deploying detailed chemical insights through CLMs yields robust results across diverse datasets and a wide range of ncAAs.

Bi-TEAM's design furthers predictive performance and generalizability by harmonizing these distinct domain knowledge bases. In the task of permeability prediction for modified peptides, Bi-TEAM achieves a 66\% increase in MCC, improves PTM druggability prediction by up to 18\%, and achieves a remarkable 350\% increase in MCC for hemolysis prediction in natural peptides. Beyond prediction, when deployed as an attribute-guidance oracle within a generative framework, Bi-TEAM boosts the success rate of generating valid cell-penetrating cyclic peptides by nearly four-fold, underscoring its potential for next-generation models that must navigate increasingly complex chemical modifications within biological scaffolds.

The bi-gated mechanism and residual fusion layer significantly enhance interpretability by revealing how local chemical modifications modulate the biological landscape to influence critical peptide properties. Our feature-ranking analyses highlight canonical yet critical factors such as N-methylation, intramolecular hydrogen bonding, and cis/trans conformations, aligning with prior membrane permeability studies \cite{rezai2006conformational, li2023cycpeptmpdb,bhardwaj2022accurate}. These findings not only confirm well-established design principles but also shed fresh light on emerging chemical modifications and their relative impact on transmembrane permeability.

Nonetheless, several limitations persist in the current approach. Creating a universal mapping from ncAAs to their closest natural analogs can prove challenging when non-canonical residues significantly diverge from the standard biological subspace. Although Bi-TEAM alleviates the token-proliferation problem by relying on the token mapping module, this approach may miss subtle nuances in cases involving entirely novel chemical moieties with no close natural counterparts, effectively creating a semantic gap between the biological and chemical representations.

Future work can focus on developing adaptive mapping mechanisms for new ncAA substituents and exploring multi-task learning approaches to handle related peptide properties such as stability, immunogenicity, or target affinity through shared parameter spaces. Additionally, integrating three-dimensional structural information could further enhance the model's ability to capture conformational effects of chemical modifications. As novel chemical designs continue to shape the boundaries of peptide therapeutics, bridging chemical and biological representation spaces through integrated computational pipelines remains essential for advancing the field. We anticipate that approaches akin to Bi-TEAM will spur further discovery of functional, stable, and bioavailable peptide drugs, ultimately expanding opportunities in peptide-focused therapeutics and translational research across diverse disease areas. Beyond peptides, this framework provides a principled pathway for modeling chemically perturbed protein matter, including engineered enzymes, synthetic biology constructs, and post-translationally regulated signaling proteins.

\section{Methods}\label{method}

\subsection{Overview}

Designing a cross-scale biochemical model necessitates the simultaneous capture of global sequence context and fine-grained atomic details. This dual requirement is particularly critical when modeling modified sequences—such as modified peptides or proteins with Post-Translational Modifications (PTMs)—where the specific chemical modifications of non-canonical amino acids play a pivotal role in functional determination. To address this challenge, we propose Bi-gaTed rEsidual spAce Modification (Bi-TEAM), a novel framework that synergistically integrates a Protein Language Model (PLM) and a Chemical Language Model (CLM) into a unified latent space. Furthermore, to enhance the model's sensitivity to structural alterations, we introduce a Modification Location Prompt, which explicitly indicates the precise coordinates of non-canonical Amino Acids (ncAAs) within the sequence (see Figure~\ref{framework}(b)). Specifically, the Bi-TEAM architecture comprises four core modules: (i) Token Mapping, designed to maintain a fixed vocabulary size while accommodating diverse inputs; (ii) Representation Learning, responsible for extracting high-dimensional biological and chemical features; (iii) Modification Prompt, which encodes explicit positional information regarding modifications; and (iv) Space Modification, a fusion mechanism that integrates and refines these cross-modal representations for downstream tasks.

\subsection{Token Mapping}
\paragraph{Biological Space Tokenization}

To integrate ncAAs into our PLM pipeline without inflating the standard protein vocabulary, we map each ncAA to its most similar canonical amino acid based on molecular similarity analysis. Only mappings exceeding a predetermined similarity threshold are retained; otherwise, the ncAA is labeled as ``$\mathbf{X}$''. Formally, we denote the complete mapping process by
\begin{equation}
\mathbf{P}_{A} = f_{\mathrm{map}} \Bigl( \mathbf{P}_{CA}, \mathbf{E}_{CA}, \mathbf{P}_{NA}, \mathbf{E}_{NA} \Bigr)
\label{eq:pa}
\end{equation}
where $\mathbf{P}_{A} = \{ \mathbf{p}_{a}^1, \mathbf{p}_{a}^2, \ldots, \mathbf{p}_{a}^{21} \}$ represents the entire set of mapped canonical amino acids for all ncAAs in a single peptide sequence. $\mathbf{P}_{CA}$ and $\mathbf{E}_{CA}$ denote the set of canonical amino acids and their corresponding Morgan fingerprints, respectively; $\mathbf{P}_{NA}$ and $\mathbf{E}_{NA}$ denote the set of all ncAAs to be mapped and their corresponding Morgan fingerprints, respectively. The detailed biological token mapping process is described in Appendix~\ref{appendix:Biotokenmapping}.

This approach ensures that non-canonical residues are replaced by only the most structurally relevant canonical amino acids, preserving evolutionary signals and structural information gleaned from large protein corpora while avoiding the addition of underrepresented or anomalous tokens.

\paragraph{Chemical Space Tokenization} 
For the CLM pipeline, we employ SELFIES \cite{krenn2020self}, a string-based method guaranteeing 100\% validity in molecular graph encoding. We convert each peptide, encompassing both canonical and non-canonical amino acids, into SELFIES-based tokens representing atoms, branches, ring notations, and bond specifications. This discretized, chemically informed representation is then fed into CLM, enabling thorough exploration of the molecular design space without relinquishing chemical validity. SELFIES's ability to generate chemically valid and diverse structures supports robust chemical space exploration, allowing Bi-TEAM to accurately learn structure--property relationships for peptides incorporating non-canonical modifications.

\subsection{Representation Learning}
Representation Learning in Bi-TEAM processes peptide input through two parallel information streams---biological encoding and chemical encoding.

\paragraph{PLM Representation} 
After obtaining the natural sequence $\textbf{S}_{p} = \{ \textbf{p}_{1}, \textbf{p}_{2}, \ldots, \textbf{p}_{n} \}$ that corresponds to non-canonical amino acids, where each $\textbf{p}_i \in \mathbf{P}_A$ represents an amino acid from the entired set of mapped canonical amino acids vocabulary, we treat this sequence as a canonical sequence and feed it into a PLM to capture evolutionary and contextual information as follows:
\begin{equation}
    \mathbf{r}_p = f_\mathrm{PLM}(\textbf{S}_p)[0]
\end{equation}
\noindent where $f_\mathrm{PLM}$ represents the PLM and we extract the `CLS' token as the representation of the entire sequence. In this work, we employ the ESM family of models, which provides high-quality sequence-level embeddings. These embeddings are subsequently used for downstream predictive models. Even in cases where a residue is labeled as ``X", ESM's ability to contextualize tokens within the surrounding sequence enables robust embedding generation.

\paragraph{CLM Representation} 
We adopt a chemical language model $f_{\mathrm{CLM}}$, which takes as input a set of SELFIES tokens
$\textbf{S}_{c} = \bigl\{ \mathbf{c}_{1}, \mathbf{c}_{2}, \ldots, \mathbf{c}_{m} \bigr\}$, where each $\textbf{c}_i \in \mathbf{C}_{NA}$. After obtaining the CLM embedding, we perform a mean pooling on it:
\begin{equation}    
\mathbf{r}_c = \mathrm{MeanPool}\bigl(f_{\mathrm{CLM}}\bigl(\mathbf{S}_{c}\bigr)\bigr)
\end{equation}
We implement $f_{\mathrm{CLM}}$ using SELF-BART~\cite{priyadarsini2024self}, a transformer-based encoder--decoder architecture inspired by BART\cite{lewis2019bart}. The model is pre-trained on large-scale molecular corpora, where randomly corrupted SELFIES sequences (via masking or permutation) are reconstructed during training. This denoising objective encourages SELF-BART to learn representations that capture high-level features such as ring strain, hydrogen-bond donor/acceptor patterns, and stereochemical configurations. 

Each input molecule is tokenized into SELFIES form. These tokens are then passed through SELF-BART's bidirectional encoder to obtain atom- and bond-level contextualized embeddings. This process distills both local (e.g., functional group) and global (e.g., conjugation, branching) molecular information, facilitating tasks such as property prediction or reactivity analysis.

The integration of PLM for biological context with CLM for chemical context establishes the foundation for constructing a unified biochemical space, which leverages established biological knowledge while incorporating the detailed chemical signals crucial for accurately modeling non-canonical amino acids within complex peptide environments.

\subsection{Modification Location Prompt}
\label{sec:prompt}

Although both the PLM and CLM provide rich embeddings for amino acid composition and molecular structure, the precise location of ncAAs in the peptide sequence can substantially influence predictive accuracy. To address this, we introduce a specifically designed ``modification location prompt" that highlights ncAA sites and integrates their functional context.

Each residue in the input sequence—whether it is canonical or non-canonical—receives a binary label (0/1) to indicate the presence of an ncAA at that position. The final output is a prompt representation $\mathbf{r}_l$, which encodes position-specific importance signals. The prompt encoder identifies critical locations and enhances their significance, allowing downstream fusion layers to focus more attention on these structurally or functionally important regions. This systematic inclusion of ncAA-position information ensures Bi-TEAM can adapt its inference process dynamically, reflecting the impact of specific substitutions on overall peptide function.

 These labels are embedded into a 256-dimensional learnable space, ensuring that the embedding vector for each token conveys whether that residue is a non-canonical amino acid. Additionally, we apply positional encoding to preserve both local and global ordering, allowing the model to distinguish residue indices across the sequence. 

We then feed the sum of the binary-label embeddings and positional embeddings into a one-layer transformer encoder (the prompt encoder), whose multi-head attention mechanism captures interdependencies among labeled and unlabeled sites.

\subsection{Bi-Gated Residual Fusion}
\label{sec:bigated}
We implement space modification via Bi-Gated residuAl fusIoN (Bi-GAIN) inspired by previous work\cite{vo2019composing, gu2021image, zhang2025sagephos}, which merges the biological-space embedding $\mathbf{r}_p$, the chemical-space embedding $\mathbf{r}_c$ and modification prompt embedding $\mathbf{r}_l$ into a unified representation, as illustrated in Fig.~\ref{framework}(b). This integration is accomplished through two parallel modules: (1) a bi-gated network that adaptively merges the protein and chemical domains, and (2) a residual layer that retrieves and integrates interdependent signals. This architecture elucidates how ncAA modifications influence critical properties by establishing connections between local chemical modifications and macroscopic functional behaviors. Both submodules simultaneously incorporate a modification location prompt embedding $\mathbf{r}_l$ to precisely specify the position of non-canonical amino acids within the sequence. The final fused embedding is obtained by summing the outputs of these two complementary modules:
\begin{equation}
\mathbf{r}_{m} 
= \underbrace{f_{\mathrm{Bi\_Gated}}(\mathbf{r}_p,\mathbf{r}_c,\mathbf{r}_l)}_{\mathbf{r}_g}
\;+\;
\underbrace{f_{\mathrm{Res}}(\mathbf{r}_p,\mathbf{r}_c,\mathbf{r}_l)}_{\mathbf{r}_r}
\label{eq:bigated_fusion}
\end{equation}

In the Bi-Gated module, each domain (biological or chemical) acts as a primary input that selectively incorporates signals from its counterpart. This architecture preserves the intrinsic features of the primary representation while assimilating complementary cues, modulated by location-specific prompts.

This residual module enriches the fused representation by synthesizing interdependent signals across all three embeddings. By bridging local chemical modifications with global peptide contexts, it captures how subtle ncAA alterations drive broader functional shifts—such as inducing structural rearrangements or modulating long-range evolutionary interactions.

The final fused embedding provides a comprehensive view of both domain-specific features and system-level interactions, thereby establishing a robust foundation for property prediction.

\subsection{Downstream Prediction}
\label{sec:prediction}

To predict diverse functional properties spanning different domains, we employ a task-specific predictor tailored to the fused embeddings. Specifically, the final representation $\mathbf{r}_m$ is fed into a three-layer Multi-Layer Perceptron (MLP) to generate classification logits. The model is optimized using the cross-entropy loss function $\mathcal{L}$, defined as:

\begin{equation}
\mathcal{L} = \mathrm{CrossEntropy}(y, \text{MLP}(\mathbf{r}_m) )
\end{equation}

where $y$ denotes the ground truth label. Further details regarding the calculation metrics are provided in Appendix \ref{metric}.

\subsection{Attribute Conditioning and Steering}
The core paradigm of existing protein design models involves converting structure prediction models into generative frameworks. Two primary strategies dominate this field: one involves fine-tuning structure prediction models as diffusion models to generate novel structures\cite{watson2023novo,krishna2024generalized}, while the alternative employs existing models without additional training by applying backpropagation to iteratively refine sequences until high model confidence scores are achieved\cite{pacesa2024bindcraft,cho2025boltzdesign1}. However, these generative models often exhibit inherent biases, such as a propensity for generating proteins rich in alpha-helices while struggling to sample beta-sheet structures\cite{geffner2025proteina}. To guide these models toward generating proteins with specific desired attributes, we incorporate gradients from the Bi-TEAM classifier into a hallucination-based generative model. 

More specifically, our objective is to design cell-penetrating cyclic peptides capable of binding the anti-VEGF drug Aflibercept, thereby enabling non-invasive ocular drug delivery\cite{fan2025peptide}. To achieve this, we first trained a Bi-TEAM model for cell-penetrating probability prediction, utilizing the dataset from \cite{kumar2025plm4cpps} and the prediction methodology outlined in Section \ref{sec:prediction}. Subsequently, to facilitate cyclic peptide generation within the BoltzDesign1 framework, we incorporated the cyclic offset as described in AfCycDesign \cite{rettie2025accurate} into the positional encoding module on the cyclic chain. To facilitate cyclic binder hallucination, we followed a four-stage optimization strategy adopted by BoltzDesign1\cite{cho2025boltzdesign1} that transitions from a relaxed, continuous sequence space to a hard, discrete representation.

\captionsetup[subfigure]{skip=2pt} %

\backmatter

\section*{Declarations}
\subsection*{Author Contributions Statement}
C.G. and Z.G. conceived the idea, developed the theoretical formalism, constructed the model and conducted experiments, and wrote the manuscript; M.H. and X.W. prepared the data; J.Z. investigated the prediction and generation developments, verified the analytical methods; H.W., Z.L., and H.C. helped write and revise the manuscript; H.C., J.B., C.H. and P.H. supervised the project and revised the manuscript.  All authors discussed the results and contributed to the final manuscript.

\subsection*{Competing Interests Statement}

The authors declare no competing interests.

\subsection*{Data and Code Availability}
This study leverages a comprehensive suite of ten datasets spanning three distinct biological domains: modified peptides, post-translational modifications (PTMs), and natural proteins/peptides.

\noindent \textbf{Modified Peptide Domain:} For membrane permeability prediction, we utilized four specific datasets: 
(1) ProPAMPA, accessible via Zenodo at \url{https://zenodo.org/records/10708332}; 
(2) ProCacoPAMPA, available from the supplementary materials of the original study at \url{https://www.cell.com/cms/10.1016/j.cell.2022.07.019/attachment/04fbb571-6234-4063-8594-7751a10b17dc/mmc7.zip}; 
(3) CycPeptMPDB v1.2, sourced from the official repository at \url{http://cycpeptmpdb.com/download/}; and 
(4) the Rezai dataset, for which SMILES representations were computationally generated using the Seq2Struc tool (\url{https://www.biosino.org/iMAC/cyclicpepedia/seq2stru}) based on the 11 data points detailed in Ref.~\cite{rezai2006conformational} (\url{https://pubs.acs.org/doi/10.1021/ja063076p}). 
For cell-penetrating peptide (CPP) prediction, we utilized the benchmark datasets established by pLM4CPPs~\cite{kumar2025plm4cpps}, available at \url{https://github.com/drkumarnandan/pLM4CPPs}.

\noindent \textbf{PTM Domain:} Datasets focusing on druggability and disease association predictions~\cite{peng2024ptm,peng2025ptm} are available via \url{https://drive.google.com/drive/folders/1RGoTzT5v_aymQQIU32IdU1-ue9Fea9la?usp=sharing}.

\noindent \textbf{Natural Protein and Peptide Domain:} Data for the three property prediction tasks—Hemolysis, Nonfouling, and Solubility—were obtained from the Peptide Dashboard repository at \url{https://github.com/ur-whitelab/peptide-dashboard/tree/master/ml/data}.

\section*{Code availability}

To ensure consistency across domains, natural peptide sequences were converted to SMILES format using RDKit. Subsequently, all SMILES strings were transformed into SELFIES representations using the RDKit and selfies Python packages. All preprocessed datasets, along with the complete source code for the Bi-TEAM framework, are publicly available at \url{https://github.com/zjgao02/Bi-TEAM}.

\begin{appendices}

\section{Dataset Description and Data Preparation}\label{dataPreparation}

\begin{figure}[htbp]
    \centering
    \includegraphics[width=0.95\textwidth,height=0.8\textheight]{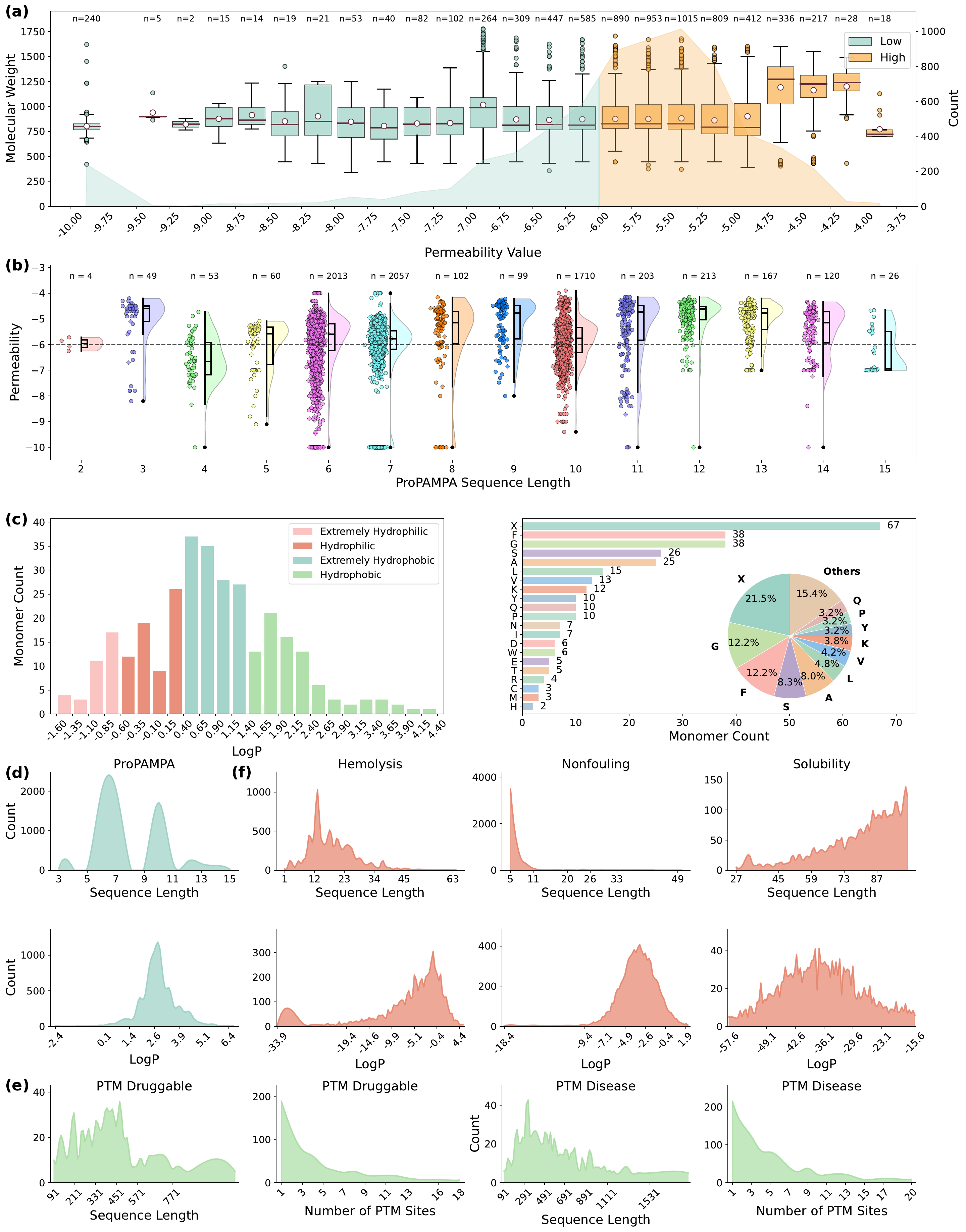}
    \caption{\textbf{Different training data distributions.}
    \textbf{(a)} Box plots illustrate the relationship between permeability and molecular weight, while the line chart shows permeability counts at 0.25 intervals, with the top indicating sample counts; the green portion on the left represents low permeability (negative), and the right represents high permeability (positive);
    \textbf{(b)} Displays the permeability distribution across different sequence lengths, with the top indicating sample counts for each length;
    \textbf{(c)} The left panel shows the LogP distribution for non-canonical monomers at 0.25 intervals, while the right panel shows the count of monomers corresponding to natural amino acids;
    \textbf{(d,e)} Sequence length and LogP distributions for modified and natural peptide datasets. 
    Panel (d) shows the modified ProPAMPA dataset, and panel (e) shows natural peptide datasets (hemolysis, nonfouling, and solubility).
    \textbf{(f)} Distributions of sequence counts and numbers of PTM sites in the two PTM datasets.}
    \label{data_analysis}
\end{figure}

To provide intuitive insights into the characteristics of our training datasets, Fig.~\ref{data_analysis} presents a multi-faceted analysis of data across three domains: modified peptides (ProPAMPA), post-translational modifications (PTM) (druggability and disease datasets), and natural peptides and proteins (Hemolysis, Nonfouling and Solubility). These visualizations illustrate key features—including sequence length distributions, PTM site distributions, amino acid frequencies, and physicochemical property distributions (such as hydrophobicity and molecular weight)—to which our model was exposed during training.  Additionally, this section introduces three modified peptides datasets: ProCacoPAMPA, CycPeptMPDB v1.2, and Rezai, which are directly used for inference based on models trained on the ProPAMPA dataset.

\subsection{Modified Peptide Datasets}

\textbf{ProPAMPA.} The ProPAMPA dataset, derived from the Cyclic Peptide Membrane Permeability Database \citep{li2023cycpeptmpdb} and curated by Geylan \textit{et al.} \citep{geylan2024methodology}, comprises 6,876 non-conjugated cyclic peptides with standardized PAMPA assay results. After removing duplicates and conflicting entries using RDKit, each peptide was classified as permeable (log~$\mathrm{P}_{\mathrm{exp}} > -6$) or non-permeable. As shown in Fig.\ref{data_analysis}(a), the dataset exhibits a class imbalance, with 67.6\% of entries labeled as permeable. Cycle sizes range from 12 to 46 atoms, sourced from multiple heterogeneous origins; some provide structurally diverse peptides, while others offer stereochemical variants of similar sequences. Fig.\ref{data_analysis}(a) utilizes box plots to depict the relationship between permeability and molecular weight, showing relatively consistent molecular weights across both classes. Fig.\ref{data_analysis}(b) employs a raincloud plot to illustrate sequence-length distributions, revealing near-normal distributions for most lengths, though with potential long tails for very short or very long sequences. Fig.\ref{data_analysis}(c) summarizes the natural and non-canonical amino acids present, highlighting the dataset's chemical diversity and the relationship between hydrophobicity and LogP values. Furthermore, Fig.~\ref{data_analysis}(d) visualizes sequence-length distributions, clarifying how chain-length variations might influence peptide properties, while also displaying the distribution of LogP values (octanol--water partition coefficients) to indicate how lipophilicity is distributed across the dataset.

\textbf{ProCacoPAMPA \cite{bhardwaj2022accurate}.} Following the data selection strategy outlined in MuCoCP \cite{yu2024mucocp}, we curated the complete collection of membrane-traversing cyclic peptides of lengths six and ten from existing work \cite{bhardwaj2022accurate}. This curated set, referred to as ``ProCacoPAMPA,'' is utilized to validate the generalization capability of the proposed algorithm.

\textbf{CycPeptMPDB v1.2 \cite{li2023cycpeptmpdb}.} This is the latest release of the largest publicly accessible database focusing on the membrane permeability of non-canonical cyclic peptides. It encompasses 8,466 records (including duplicates) compiled from 56 publications. In addition to providing a mapping between non-canonical monomers and their corresponding natural amino acids, each peptide entry includes detailed sequence information highlighting the presence of non-canonical monomers, along with chemical SMILES that can be readily converted to SELFIES. In this study, we removed duplicate entries found in the ProPAMPA dataset, resulting in a refined subset of 1,230 data points.

\textbf{Rezai \cite{rezai2006conformational}.} This dataset contains passive membrane permeability measurements for 11 cyclic peptides derived from PAMPA assays. Notably, the transmembrane permeability values in this dataset are all lower than $-6$, constituting an exclusively negative set of data points.

\subsection{PTM Datasets}

\textbf{PTM Druggability Dataset \cite{peng2024ptm,peng2025ptm}.}
The druggability prediction data is used to evaluate PTM sequences that affect therapeutic targets, specifically focusing on how modifications alter protein structure and the accessibility of binding sites \cite{li2022dbptm}. The first two columns of Fig.~\ref{data_analysis}(e) present the statistics for sequence length and PTM sites within the druggability dataset. These visualizations reveal that the dataset is characterized by significant sequence lengths and a long-tailed distribution of PTM sites, presenting a substantial challenge for prediction tasks.

\textbf{PTM Disease Dataset \cite{peng2024ptm, peng2025ptm}.}
The disease association data is curated from the dbPTM database \cite{li2022dbptm}, linking PTMs to diseases such as cancer, neurodegenerative disorders, and diabetes. Annotations are sourced from databases including PhosphoSitePlus, ActiveDriverDB, and Genome-Wide Association Studies (GWAS), as well as manual curation \cite{hornbeck2012phosphositeplus, krassowski2021activedriverdb}. The last two columns of Fig.~\ref{data_analysis}(e) provide statistics on sequences and PTM sites for this dataset. Compared to the druggability data, the distribution of PTM sites remains similar; however, the statistical distribution of sequence lengths spans a significantly broader range.

\subsection{Natural Peptide and Protein Datasets}

\textbf{Hemolysis Dataset \cite{ansari2023serverless}.} Hemolysis refers to the destruction of red blood cell membranes, a critical factor in the safe application of antimicrobial peptides (AMPs). This dataset, derived from the DBAASPv3 database \cite{gogoladze2014dbaasp}, is used to predict peptide hemolytic activity. Peptides with activity below 100~$\mu$g/mL are classified as hemolytic \cite{ansari2023serverless}. Each measurement is treated as an independent case, implying that the same sequence may appear multiple times. The training set contains 9,316 sequences, of which 19.6\% are positive (hemolytic) and 80.4\% are negative (non-hemolytic). These sequences consist solely of L-form and canonical amino acids. Notably, about 40\% of the observations include identical sequences labeled as both negative and positive, highlighting the complexity of distinguishing hemolytic from non-hemolytic peptides \cite{guntuboina2023peptidebert}. Fig.\ref{data_analysis}(f) further characterizes this dataset through different analytical perspectives, showing count distributions regarding sequence length and LogP. The upper row of Fig.\ref{data_analysis}(f) demonstrates a wide range of peptide lengths, indicating significant structural diversity within the dataset. The lower row of Fig.\ref{data_analysis}(f) reveals a more complex LogP distribution compared to the other datasets, suggesting greater variability in the hydrophobic properties of these peptides.

\textbf{Nonfouling Dataset \cite{barrett2018classifying, ansari2023serverless}.} This dataset comprises short peptide sequences designed to resist unwanted adsorption \cite{guntuboina2023peptidebert}. These peptides are pivotal in enhancing the biocompatibility, durability, and overall performance of engineered biomaterials, medical devices, and drug delivery systems. The second column of Fig.~\ref{data_analysis}(f) demonstrates that sequence lengths in this dataset primarily range between 5 and 10 residues. The LogP distribution follows a normal distribution, and visualization shows samples clustering well together, suggesting that algorithms are likely to perform effectively on this dataset. Comparative experiments confirmed these observations.

\textbf{Solubility Dataset \cite{ansari2023serverless}.} This dataset comprises protein sequences annotated via PROSO II \cite{smialowski2012proso}, with solubility assignments derived from retrospective analyses under the Protein Structure Initiative. Notably, the final two columns of Fig.~\ref{data_analysis}(f) display the statistical distributions of protein sequences and LogP properties for this dataset.

\section{Evaluation Metric Description}\label{metric}
To comprehensively assess model performance in predicting diverse properties, we evaluate our approach primarily using six metrics. Each metric offers a unique perspective on classification results, ensuring a thorough and balanced analysis across diverse usage scenarios and property classes.
\begin{enumerate}

    \item \textbf{Matthews Correlation Coefficient (MCC)}
    \begin{equation}
    \mathrm{MCC} \;=\; \frac{(\mathrm{TP} \times \mathrm{TN})\;-\;(\mathrm{FP} \times \mathrm{FN})}{\sqrt{(\mathrm{TP}+\mathrm{FP})\,(\mathrm{TP}+\mathrm{FN})\,(\mathrm{TN}+\mathrm{FP})\,(\mathrm{TN}+\mathrm{FN})}}
    \label{eq:mcc}
    \end{equation}
    MCC \cite{matthews1975comparison,baldi2000assessing} combines all elements of the confusion matrix: true positives (TP), true negatives (TN), false positives (FP), and false negatives (FN). Its range spans \(-1\) to \(+1\), where \(+1\) indicates perfect classification, \(0\) denotes random guessing, and \(-1\) indicates a complete mismatch. MCC is particularly resilient to class imbalance, making it a primary metric in our evaluation.

    \item \textbf{Accuracy}
    \begin{equation}
    \mathrm{Accuracy} \:=\: \frac{\mathrm{TP} + \mathrm{TN}}{\mathrm{TP} + \mathrm{TN} + \mathrm{FP} + \mathrm{FN}}
    \label{eq:accuracy}
    \end{equation}
    Accuracy measures the proportion of correctly classified instances among all instances. Although it offers a straightforward overview, it can be misleading with highly imbalanced data.

    \item \textbf{Area Under the Receiver Operating Characteristic Curve (ROC-AUC)}
    ROC-AUC evaluates a model's ability to discriminate between classes at various threshold settings. The ROC curve plots the true positive rate against the false positive rate. The area under this curve varies from 0.5 (random guessing) to 1.0 (perfect separation), making ROC-AUC a threshold-agnostic measure of diagnostic power.

    \item \textbf{Precision}
        \begin{equation}
        \mathrm{Precision}  \:=\: \frac{\mathrm{TP}}{\mathrm{TP} + \mathrm{FP}}
        \label{eq:precision}
        \end{equation}
    Precision represents the proportion of true positive results among all positive predictions made by a model. High precision indicates that the model effectively minimizes false positives, giving users greater confidence in positive predictions. 
    \item \textbf{F1-score}
    \begin{equation}
    \mathrm{F1 \textendash score} = 2 \times \frac{\mathrm{Precision} \times \mathrm{Recall}}{\mathrm{Precision} + \mathrm{Recall}}
    \label{eq:f1-score}
    \end{equation}
    As the harmonic mean of precision and recall ($\mathrm{r
    ecall} = \frac{\mathrm{TP}}{\mathrm{TP} + \mathrm{FN}}$), the F1-score maintains a balance between avoiding false negatives and false positives. It is especially helpful when the dataset is imbalanced, reflecting both misclassification risks.

    \item \textbf{CrossEntropy}
    \begin{equation}
    \mathrm{CrossEntropy} \:=\: -\,\bigl[\, y\,\log(p) \;+\; (1 - y)\,\log(1 - p) \bigr]
    \label{eq:cross_entropy}
    \end{equation}
    $y$ represents the ground-truth label (0 or 1) and $p$ is the predicted probability for belonging to the positive class. This function is commonly referred to as log loss. It measures how closely the predicted probability distribution matches the actual distribution. A lower cross entropy indicates that predictions align well with reality.
\end{enumerate}
By jointly considering these metrics, our analysis elucidates not only whether the model correctly classifies samples, but also how balanced and stable its predictions are across varied application scenarios and property classes.

\section{In-Depth Performance Analysis Through Confusion Matrix}\label{ConfusionMatrix}

\begin{figure}[htbp!]
    \centering
\includegraphics[width=1.0\textwidth]{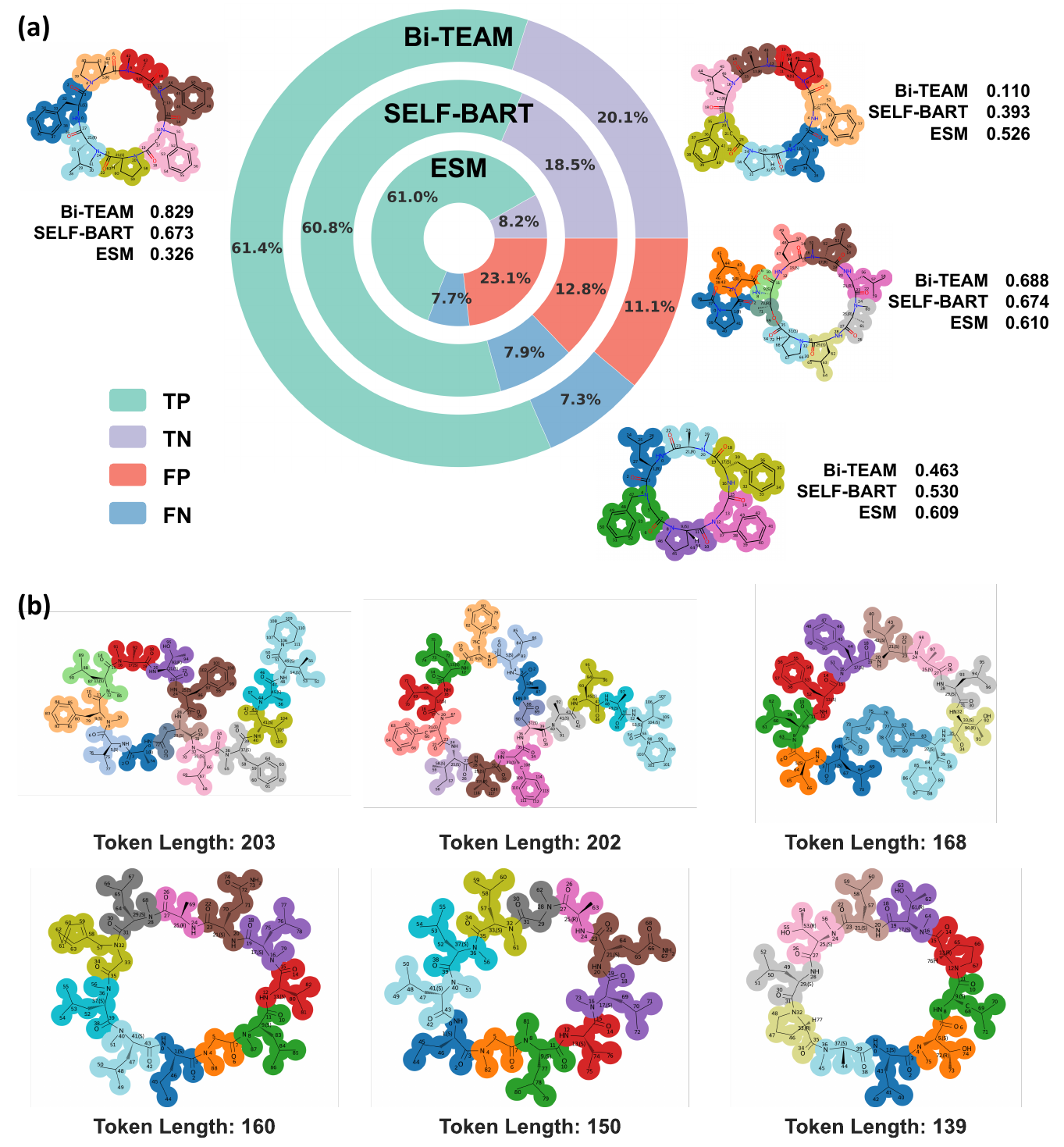} 
    \caption{\textbf{Comprehensive evaluation through confusion matrix.} \textbf{(a)} Each sample represents the logits of each model predicting that it can pass through the cell membrane. Logits $\geq 0.5$ means that it is predicted to be permeable, and $< 0.5$ means otherwise. \textbf{(b)} Visualization of six representative cases along with their tokenized lengths. SELF-BART model produces incorrect predictions for these examples, the other models correctly identify the intended outputs.}
    \label{confusion}
\end{figure}

Using the 1375 test samples from ProPAMPA, we conducted a detailed confusion matrix analysis on the prediction preferences of different base models. As evident in \ref{confusion}(a), Bi-TEAM clearly achieves the optimal balance of True Positives (TP) and True Negatives (TN) while minimizing False Positives (FP) and False Negatives (FN). By contrast, although ESM also shows a relatively high TP rate (indicating that it recognizes some beneficial sequence-level signals learned from protein evolution), it suffers from a notably higher FP rate. This discrepancy likely arises because ESM's amino acid–level embeddings do not capture site-specific chemical modifications that deviate from the canonical residues; consequently, ESM may misinterpret certain chemically altered side chains as valid functional motifs, elevating the false-positive count.

On the other hand, SELF-BART, which encodes molecular string exclusively at the chemical level, provides improved sensitivity for non-canonical modifications compared to ESM but lacks the evolutionary nuances and peptide-scale coordination signals that govern long peptide sequences. Consequently, as illustrated in Fig. \ref{confusion}(a), SELF-BART's confusion matrix displays modest decreases in both true positives (TP) and true negatives (TN) relative to Bi-TEAM, accompanied by elevated false negatives (FN) and a slight increase in false positives (FP).

To further elucidate this limitation, we examined six instructive test-set examples where ESM and Bi-TEAM made correct predictions whereas SELF-BART erred. These cases are depicted in detail in Fig. \ref{confusion}(b). Although the ProPAMPA dataset features relatively short peptide sequences that do not exceed SELF-BART's input window, these failing test cases underscore the model's limited capacity to capture subtle chemical modifications in longer sequences and highlight an inherent vulnerability when encountering natural peptides that do surpass its context limit, forcing truncation and ultimately sacrificing critical global evolutionary signals. By contrast, ESM and Bi-TEAM integrate global evolutionary information more effectively, thus emphasizing the importance of robust global context integration for accurate peptide-level predictions.

Bi-TEAM effectively addresses the limitations of both ESM and SELF-BART through our multi-scale fusion approach that integrates sequence embeddings (capturing evolutionary and peptide-structure information) with chemical-level embeddings (capturing intricate chemical modifications).
The result is a more precise confusion matrix distribution: TP and TN are well balanced, indicating strong recognition of both correctly modified peptides and genuine negatives, while FP and FN remain low. Notably, the modification location prompt precisely highlights where non-canonical changes occur and helps the model reconcile local chemical variability with the broader biological context.
Consequently, Bi-TEAM successfully avoids ESM's blind spots with respect to non-canonical modifications and SELF-BART's inability to grasp peptide-scale evolution, yielding more robust peptide property predictions overall.

\section{Data Split Methods}\label{dataSplitMethods}
We employed widely accepted molecular fingerprinting techniques to partition datasets based on molecular similarity. Specifically, we utilized MinHashed fingerprints \cite{capecchi2020one}, a versatile approach applicable to both small molecules and peptide. For ProPAMPA and three natural peptide datasets, we generated fingerprints with a length of 2048 bits to capture structural information comprehensively. Subsequently, we applied K-means clustering with k=6 to segregate the molecular space \cite{hartigan1979algorithm}. Following dimensionality reduction via Principal Component Analysis (PCA) to two dimensions \cite{abdi2010principal}, we selected the cluster most distant from the other five clusters as our test set, maximizing the distribution shift between training and evaluation data to assess out-of-distribution generalization capability rigorously.

\section{Exploring Drug Screening Potential Through External Datasets Testing}\label{ood_wet}

\begin{figure}[htbp!]
    \centering
\includegraphics[width=1.0\textwidth]{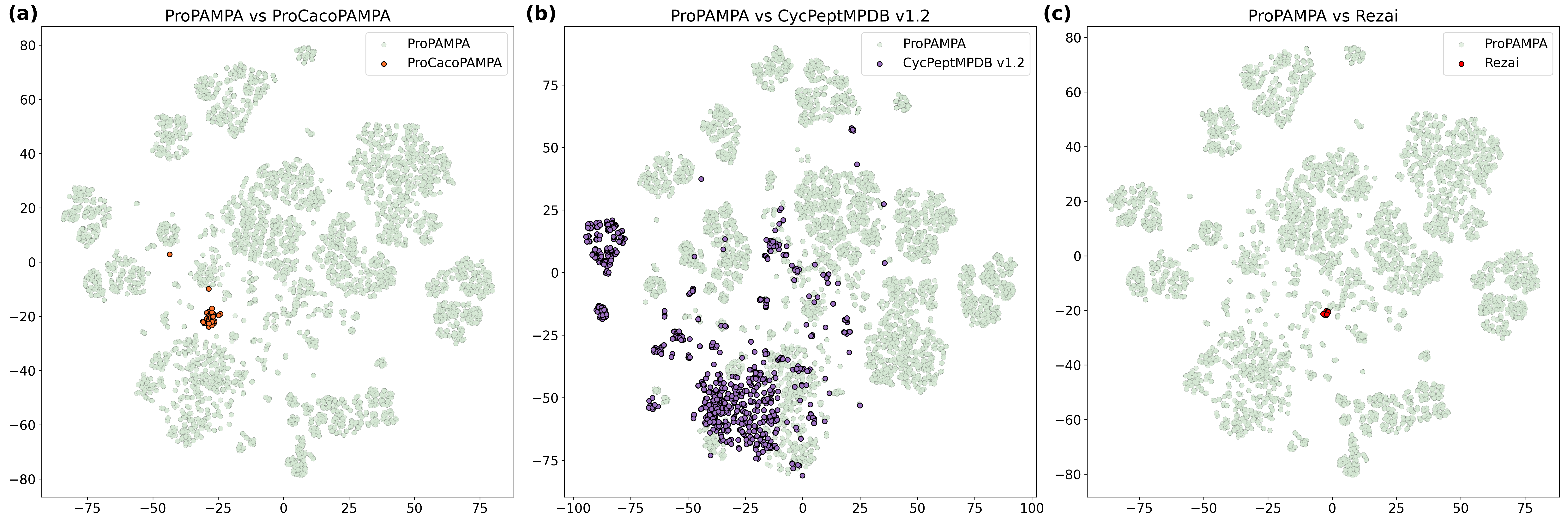} 
    \caption{\textbf{t-SNE visualization of molecular fingerprints distribution on three external datasets.} Each plot displays two labels: the ProPAMPA dataset used for training, and the respective external dataset - (a) ProCacoPAMPA, (b) CycPeptMPDB v1.2, and (c) Rezai. All these three datasets encompass molecules with diverse scaffolds and compounds with novel structural modifications. They are validated through wet lab experiments, serving as reliable benchmarks for practical applications of our method.}
    \label{SI_ood_tse}
\end{figure}

We utilized t-SNE \cite{van2008visualizing} to reduce the 2048-dimensional features to two dimensions for visualization. In Fig. \ref{SI_ood_tse}, we observe that ProCacoPAMPA forms isolated high-density clusters away from ProPAMPA, CycPeptMPDB v1.2 contains both completely out-of-distribution instances and samples within the original distribution, and Rezai test cases demonstrate some correlation with the original data distribution.

\section{Generalization Assessment on Post-Translational Modificatio}\label{SI_PTM}

\begin{figure}[htbp!]
    \centering
\includegraphics[width=1.0\textwidth]{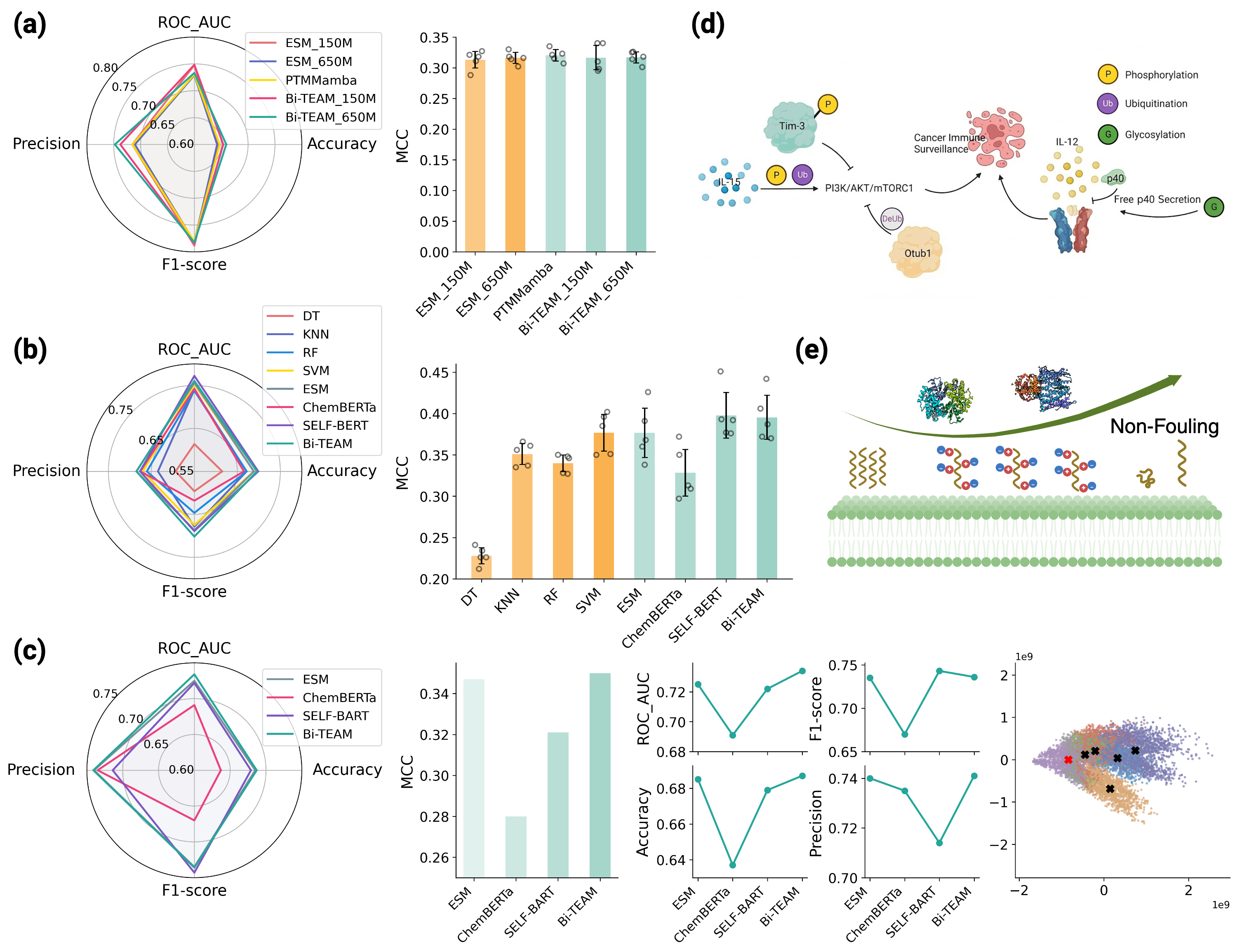} 
\caption{\textbf{Experimental results comparison across PTM disease datasets and natural non-fouling datasets using multiple metrics.} \textbf{(a)} Performance on the prediction of whether PTMs cause disease, with radar charts displaying average values across four metrics and bar charts providing detailed individual results with discrete points and error bars showing standard deviation. \textbf{(b)} Results on the Non-fouling dataset for non-fouling prediction. \textbf{(c)} Non-fouling prediction based on a principled cluster-based train-test partitioning strategy derived from fingerprint similarity distributions. \textbf{(d)} Illustrative diagram of PTM-induced disease\cite{wang2023post}. \textbf{(e)} Illustrative diagram of peptide non-fouling properties.}
    \label{SI_NaturalPTM}
\end{figure}

Building upon the PTM dataset evaluation, we extended our analysis to PTM-disease association prediction, with the underlying biological mechanism illustrated in Fig.~\ref{SI_NaturalPTM}(d). Specifically, cytokine-associated PTMs orchestrate the activation, inhibition, and secretion patterns of key signaling pathways. This fine-grained regulation reshapes the landscape of immune surveillance, thereby influencing the pathogenesis of a wide spectrum of diseases.

Fig.~\ref{SI_NaturalPTM}(a) presents a comparative performance assessment of various algorithms. In contrast to the PTM druggability prediction task, this association task proves significantly more challenging, characterized by lower overall metric scores and narrower performance margins between different methods. Despite these inherent difficulties, the proposed Bi-TEAM demonstrates the best overall performance. Notably, Bi-TEAM achieves a relative improvement of nearly 5\% in Precision compared to the strongest baseline, PTMMamba. Furthermore, it maintains state-of-the-art performance in Accuracy and ROC-AUC, while remaining highly competitive in the other two metrics.

Synthesizing these results with the aforementioned PTM-druggability analysis reveals that Bi-TEAM exhibits robust and superior performance across distinct downstream tasks. On one hand, it effectively identifies potential therapeutic targets at the PTM-drug interface; on the other, it maintains a leading position in PTM-disease association prediction. This dual capability demonstrates that the representations learned by our model possess strong cross-task generalization. Consequently, this establishes a solid methodological foundation for the simultaneous mining of "druggable PTMs" and "pathogenic PTMs" within a unified modeling framework.

\section{Generalization Assessment on Natural Peptide Dataset}\label{natural}

To further validate the versatility and robust generalization capability of our proposed Bi-TEAM on natural protein datasets, we provide a comprehensive comparison of Bi-TEAM and seven baselines on natural peptide Non-fouling across different metrics in Fig.~\ref{SI_NaturalPTM}.

% Fig.~\ref{appendix_natural}(a) presents the comparative experimental results on the natural peptide Solubility dataset. The results show that Bi-TEAM is significantly ahead in key indicators such as MCC, F1-score and Precision, achieving improvements of 8.4\%, 3.0\%, and 2.3\% respectively compared to the best baseline model ESM. Compared to deep learning approaches, traditional machine learning models (such as DT, KNN, RF, and SVM) do not come close to Bi-TEAM or ESM in Accuracy or ROC\_AUC, although they may occasionally achieve comparable Precision in certain runs. Overall, they are more sensitive to data partitioning, leading to more fluctuation in F1-score and MCC.

Fig. \ref{SI_NaturalPTM}(b) presents the results on the natural protein Non-fouling dataset. As illustrated in the visualization in the rightmost column, the sample points in this dataset are highly concentrated, enabling all evaluated methods to achieve excellent predictive performance. Consequently, the competition among algorithms is intense across all metrics. Nevertheless, Bi-TEAM distinguishes itself by improving the F1-score by nearly 2\% compared to the top-performing baseline and achieves the highest precision, tying with ESM. regarding the remaining metrics, SELF-BART and our proposed method emerge as the two most competitive approaches, exhibiting only marginal differences. Notably, traditional machine learning algorithms such as SVM, RF, and KNN perform comparably to the state-of-the-art on certain metrics within this non-fouling prediction task. This demonstrates that classical approaches remain effective on datasets characterized by low feature space complexity. Overall, this performance comparison further validates the necessity of integrating peptide sequence representations with fine-grained chemical descriptions.

Fig. \ref{SI_NaturalPTM}(c) illustrates the experimental results under this similarity-based train-test strategy. In contrast to the random-split scenario, Bi-TEAM demonstrates significantly superior performance, achieving state-of-the-art results across all metrics with the exception of the F1-score. Even regarding the F1-score, the proposed method and SELF-BART remain the most competitive approaches.

\begin{figure}[htp]
    \centering
\includegraphics[width=1.0\textwidth]{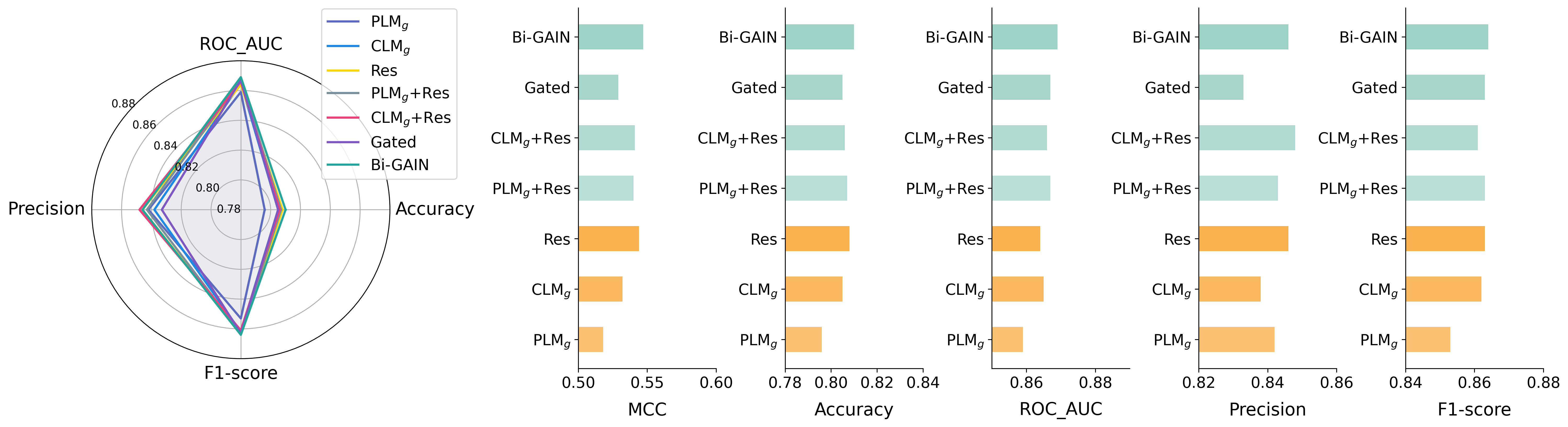} 
\caption{\textbf{Comparative performance of submodule in Bi-GAIN.}}
    \label{appendix_ablation}
\end{figure}

\section{Ablation Study of Submodules in Bi-Gated Residual Fusion}\label{appendix:ablationStudy}
To investigate the contribution of each submodule in the proposed Bi-Gated ResiduAl FusIoN (Bi-GAIN), Fig.~\ref{appendix_ablation} compares the performance of various submodules. To ensure that we accurately isolate the impact of each fusion submodule, we did not apply any prompt-based strategy. The figure presents the comprehensive performance of PLM gated, CLM gated, and Residual, as well as their various combinations, in terms of Accuracy, ROC\_AUC, Precision, F1-Score, and MCC. For simplicity, we refer to PLM gated, CLM gated, and Residual as PLM$_g$, CLM$_g$, and Res, respectively; PLM$_g$ + Res, CLM$_g$+Res, and Gated denote any pairwise combination; while Bi-GAIN is the integrated weighted result of all three submodules.

Overall, these submodules and their combinations exhibit relatively close performance across different metrics, but Bi-GAIN typically achieves stable and significant improvements under most metrics. Notably, for classification-discriminative metrics such as MCC, Accuracy, ROC\_AUC, and F1-Score, Bi-GAIN generally outperforms all other configurations or single submodules. This indicates that after integrating the features from biology (protein language model, PLM) and chemistry (chemical language model, CLM), supplemented by the Residual module's complementary signals, not only can we retain the fundamental advantages of each representation (e.g., the PLM's global sequence semantics and the CLM's fine-grained chemical details), but we can also fully capture the interdependencies between them. Consequently, the model can develop more representative multimodal features of non-canonical peptides, ultimately boosting prediction accuracy.

\section{More Cases for Prompt Effectiveness}\label{appendix:caseStudy}

\begin{figure}[htp]
    \centering
\includegraphics[width=1.0\textwidth]{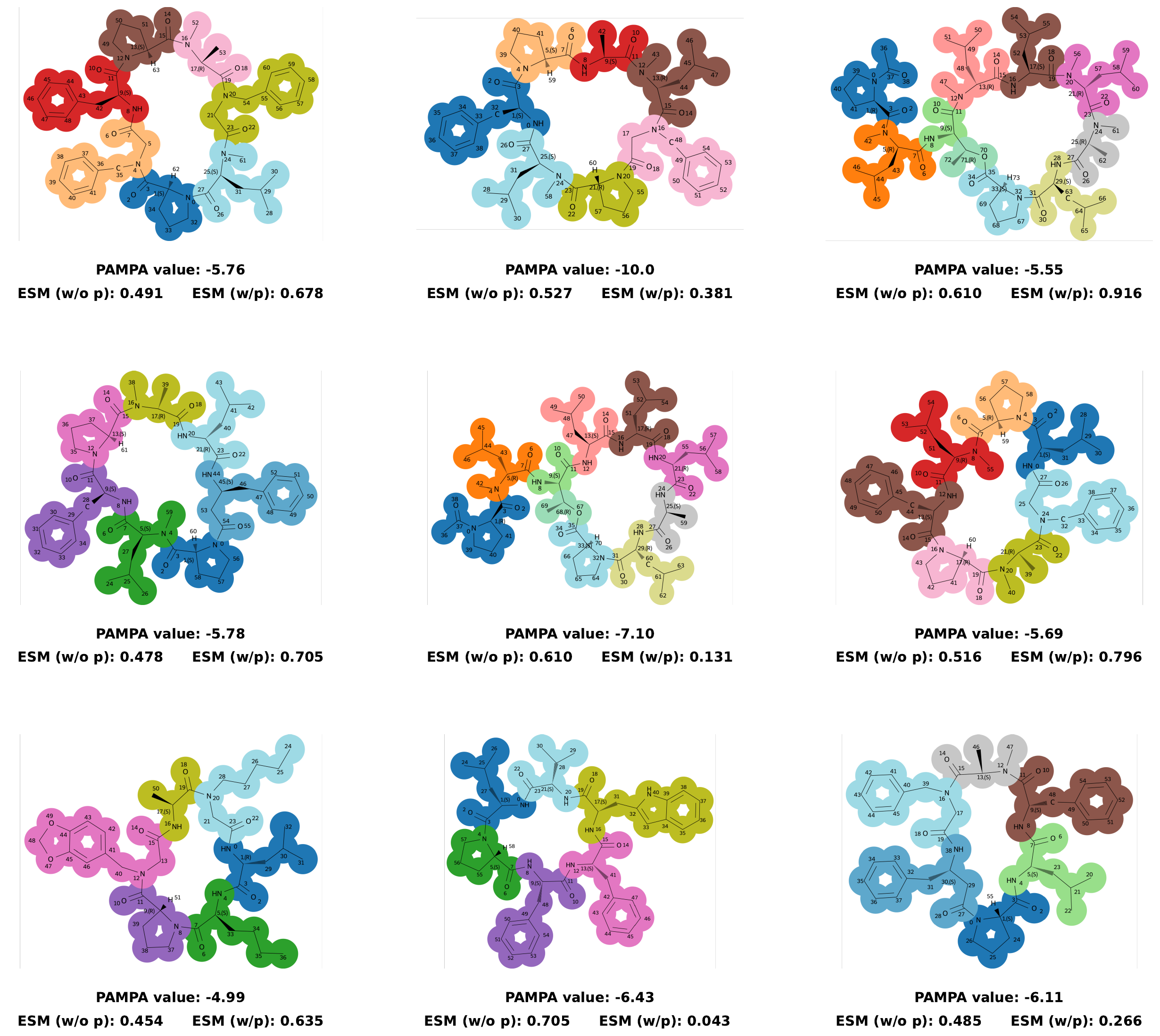} 
\caption{\textbf{Additional comparative examples of ESM model predictions with and without prompts.}}
    \label{appendix_ablation_ESM}
\end{figure}

\begin{figure}[htp]
    \centering
\includegraphics[width=1.0\textwidth]{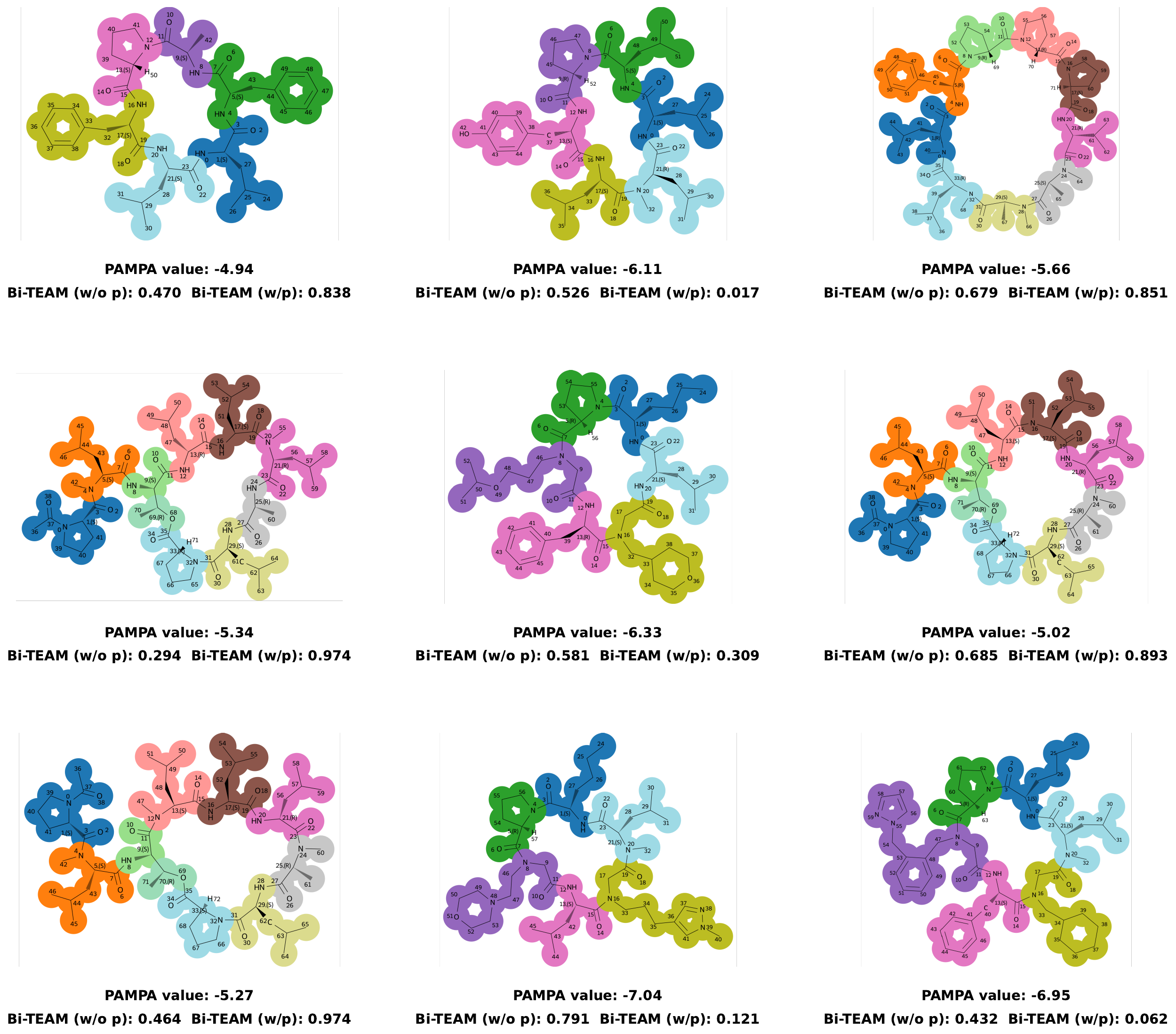} 
\caption{\textbf{Additional comparative examples of Bi-TEAM model predictions with and without prompts.}}
    \label{appendix_ablation_Ours}
\end{figure}

Fig.~\ref{appendix_ablation_ESM}~and~\ref{appendix_ablation_Ours} respectively present additional prediction improvement examples for ESM and Bi-TEAM under scenarios with and without prompts. In the figures, \textit{w/o p} and \textit{w/p} indicate models operating without and with prompts, respectively. A PAMPA value greater than -6 designates a positive sample, otherwise negative; likewise, a predicted value above 0.5 is considered positive and below 0.5 negative. The extensive comparisons in both figures further demonstrate the necessity of incorporating the Modification Location Prompt.

\section{Extended Methodology and Additional Results for Rank-Based Space Modification Analysis}\label{appendix:rank}
\begin{figure}[htbp!]
    \centering
\includegraphics[width=1.0\textwidth]{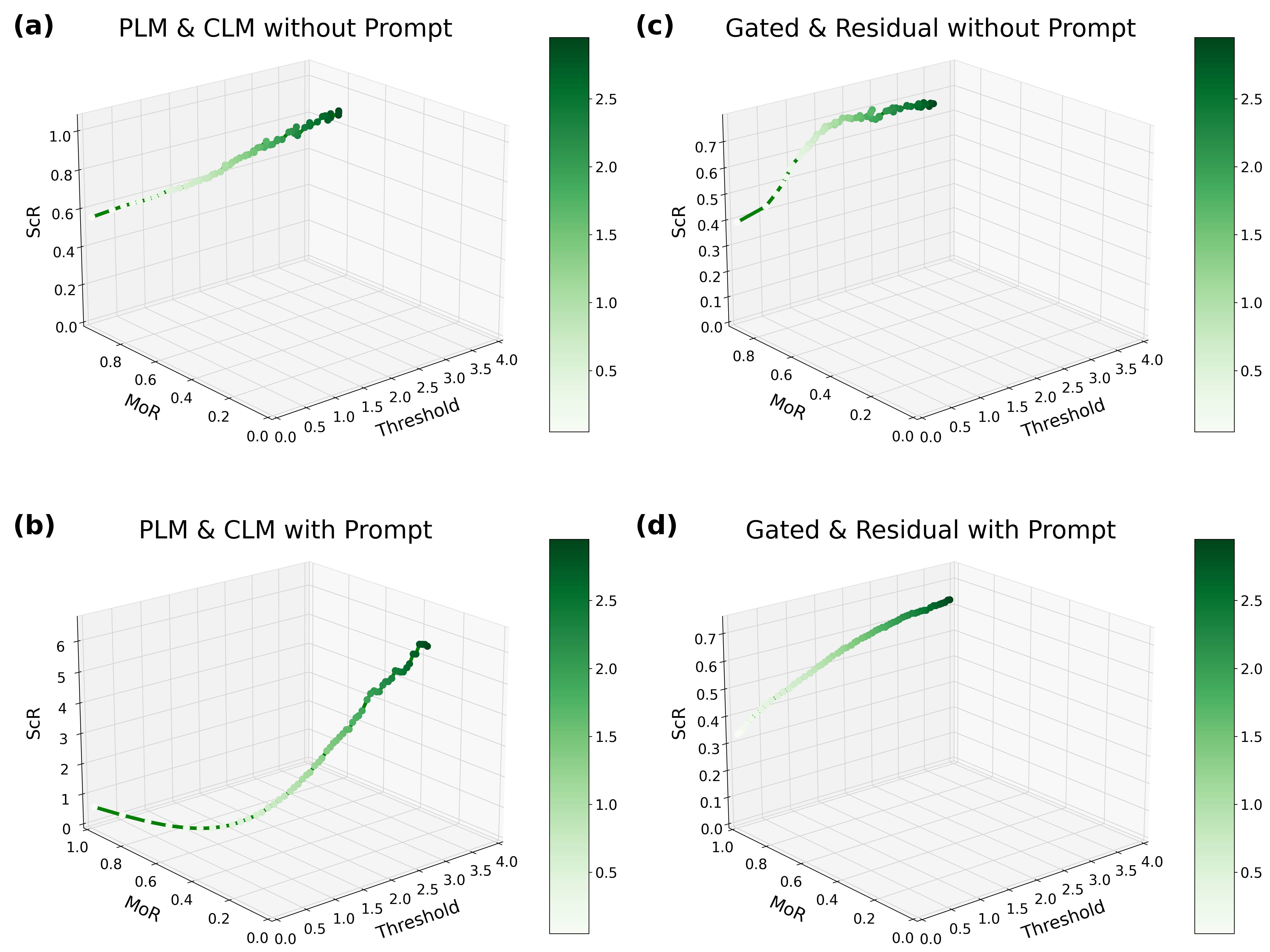} 
    \caption{\textbf{Rank Analysis.}\textbf{(a)} and \textbf{(b)} show the rank comparison between PLM, CLM, and the fusion feature in the without-prompt and with-prompt scenarios, respectively. \textbf{(c)} and \textbf{(d)} present the rank comparison among PLM gated features, CLM gated features, a residual feature, and their fusion feature under the without-prompt and with-prompt scenarios, respectively.}
    \label{rank}
\end{figure}

Rank characterizes the number of independent feature dimensions and thus serves as an effective measure of the representational capacity of an embedding. In space modification studies, data from different modalities are typically projected into a shared low-dimensional subspace, while each modality preserves its own modality-specific representation to capture information that cannot be fully expressed in the shared space \cite{gao2023long,zhang2015low}. To further elucidate this concept, we conducted an experiment examining the role of Singular Value Decomposition (SVD) in cross-modal embedding analysis \cite{wall2003singular}. First, we define two concepts that help illustrate the relationships among different ranks:

\begin{equation}
{\mathrm{MoR}}=\frac{\max \{\operatorname{r}(A_{1}),\operatorname{r}(A_{2}),..., \operatorname{r}(A_n)\}}{\operatorname{r}(F)}
\end{equation}
Here, the Max-over Ratio (MoR) indicates the ratio of the largest single-modality rank to the fused modality rank; $\operatorname{r}(A)=\{\operatorname{r}(A_{1}),\operatorname{r}(A_{2}),..., \operatorname{r}(A_n)\}$ denotes the rank values of each single modality, and $\operatorname{r}(F)$ is the rank of the fused representation.
\begin{equation}
{\mathrm{ScR}}=\frac{\operatorname{r}(F)}{\min \{\operatorname{r}(A_1) + \operatorname{r}(A_2) + ... + \operatorname{r}(A_n), \, \operatorname{r}_{\max}\}}
\end{equation}
The Sum-coverage Ratio (ScR) represents the ratio of the fused modality's rank to the sum of all single-modality ranks, and $\operatorname{r}_{\max}$ is the maximum possible rank, typically the embedding dimensionality.

We performed SVD on the embeddings of the PLM and CLM models separately, and retained only the singular values above different thresholds, thereby obtaining their ranks $\operatorname{r}(P)$ and $\operatorname{r}(C)$. We then applied the bi-gated residual network to fuse these two embedding sets into a single embedding matrix, again using SVD to estimate its rank $\operatorname{r}(F)$.

As shown in Fig.~\ref{rank}, we compare rank values in two scenarios: with and without prompts. Fig.~\ref{rank}(a) presents the rank comparison of PLM, CLM, and the fusion representation under the no-prompt condition. By setting different thresholds for the minimum effective singular value, we observe that both MoR and ScR remain within [0,1]. Empirical results reveal a consistent phenomenon:
\begin{equation}
    \max\{\operatorname{r}(P),\,\operatorname{r}(C)\} \;\;<\;\; \operatorname{r}(F) \;\;\leq\;\; \min\{\,\operatorname{r}(P) + \operatorname{r}(C),\,\operatorname{r}_{\max}\}
\end{equation}
where $\operatorname{r}_{\max}$ is the embedding dimensionality. We hypothesize this can be explained through the following observations: \textbf{1) Uncorrelated Modalities without Overlapping Information:} If two modalities contain entirely disjoint features, the fused matrix incorporates all such features, pushing its rank closer to $\operatorname{r}(A)+\operatorname{r}(B)$. \textbf{2) Strongly Correlated Modalities With Redundant Information:} When two modalities exhibit overlap in a shared subspace, the fused representation does not simply add their ranks; Overlapping features contribute redundant dimensions, causing a smaller increase in rank than the naive sum would suggest. \textbf{3) Rank Elevation by Information Gain:} Compared to a single modality, the fused representation leverages additional cross-modal interactions, often exceeding $\max\{\operatorname{r}(A),\,\operatorname{r}(B)\}$, thereby underscoring the benefits of multimodal fusion.

Next, Fig.~\ref{rank}(b) shows the rank comparison between PLM, CLM, and the fusion feature under the with-prompt scenario. Relative to no-prompt conditions, two major changes emerge: first, ScR values can exceed 1 at higher thresholds; second, MoR drops substantially overall. Both phenomena derive from the added information introduced by the modification location prompt, indicating that the fused representation extends beyond the original scope of PLM and CLM. Another interesting observation is that, in both with- and without-prompt scenarios, increasing the threshold generally raises ScR and lowers MoR. We attribute this to the exclusion of more ``noise" at higher thresholds, thereby retaining only more meaningful singular values to produce a more reliable rank estimate. Such complementary trends of ScR and MoR further affirm the capacity of our fusion method to generate high-quality representations.

Moreover, since our fused representation ultimately results from a weighted combination of PLM gated features, CLM gated features, and a residual feature, we also compare their individual ranks against the final fused rank. Fig.~\ref{rank}(c) and (d) display both no-prompt and with-prompt results, providing a perfect fit for the following inequality:

\begin{equation}
    \max\{\operatorname{r}(P_g),\,\operatorname{r}(C_g),\,\operatorname{r}(R)\} \;\;<\;\; \operatorname{r}(F) \;\;\leq\;\; \min\{\,\operatorname{r}(P_g) + \operatorname{r}(C_g) + \operatorname{r}(R),\,\operatorname{r}_{\max}\}
\end{equation}
where $\operatorname{r}(P_g)$, $\operatorname{r}(C_g)$, and $\operatorname{r}(R)$ represent the ranks of the PLM gated feature, the CLM gated feature, and the residual feature, respectively. Notably, in Fig.~\ref{framework}(c), the framework highlights three sub-networks---PLM gated, CLM gated, and Residual---each producing distinct embedding distributions. Their interactions in the Space Modification module align closely with the rank analysis presented here. From a rank perspective, these complementary subspaces seamlessly explain the mechanism behind our fusion design and further underscore their mutual contributions in the final fused representation.

\section{Training setting}\label{trainSetting}
In order to unify information from both the biological and chemical domains, we draw on two distinct Language Models without fine-tuning their parameters. For the biological space, we employ ESM2\_8M as our default Protein Language Model (PLM) \cite{lin2023evolutionary}; For the chemical space, SELF-BART \cite{priyadarsini2024self} serves as our Chemical Language Model (CLM). Because our primary focus is on integrating multi-scale representations rather than improving the LMs themselves, both the PLMs and CLM remain frozen throughout training.

After extracting each sequence's peptide-level embedding, chemical-level embedding, and the modification prompt keying ncAA locations, we feed these three streams into Bi-TEAM. The resulting fused representation is then passed through a three-layer Multilayer Perceptron (MLP). The first two layers of the MLP are followed by ReLU activation functions, while the final layer outputs class logits. We optimize this network using the cross entropy loss, with Equation \ref{eq:cross_entropy} and its explanatory content providing more details.

To ensure fairness and consistency in our experiments, we fix the learning rate at \(5\times10^{-5}\), use a maximum of 100 training epochs, set the batch size to 128, and apply a weight decay of 0.001. Throughout, we adopt the AdamW optimizer \cite{kingma2014adam} for parameter updates. All experiments were conducted on a single NVIDIA A40 GPU. These settings are maintained uniformly across different configurations of the ESM2 (8M/35M/150M/650M) and SELF-BART models, highlighting the versatility and robustness of our architecture in predicting diverse non-canonical amino acid properties.

\section{Biological Token Mapping}\label{appendix:Biotokenmapping}

\begin{figure}[h!]
    \centering
     \includegraphics[width=\textwidth]{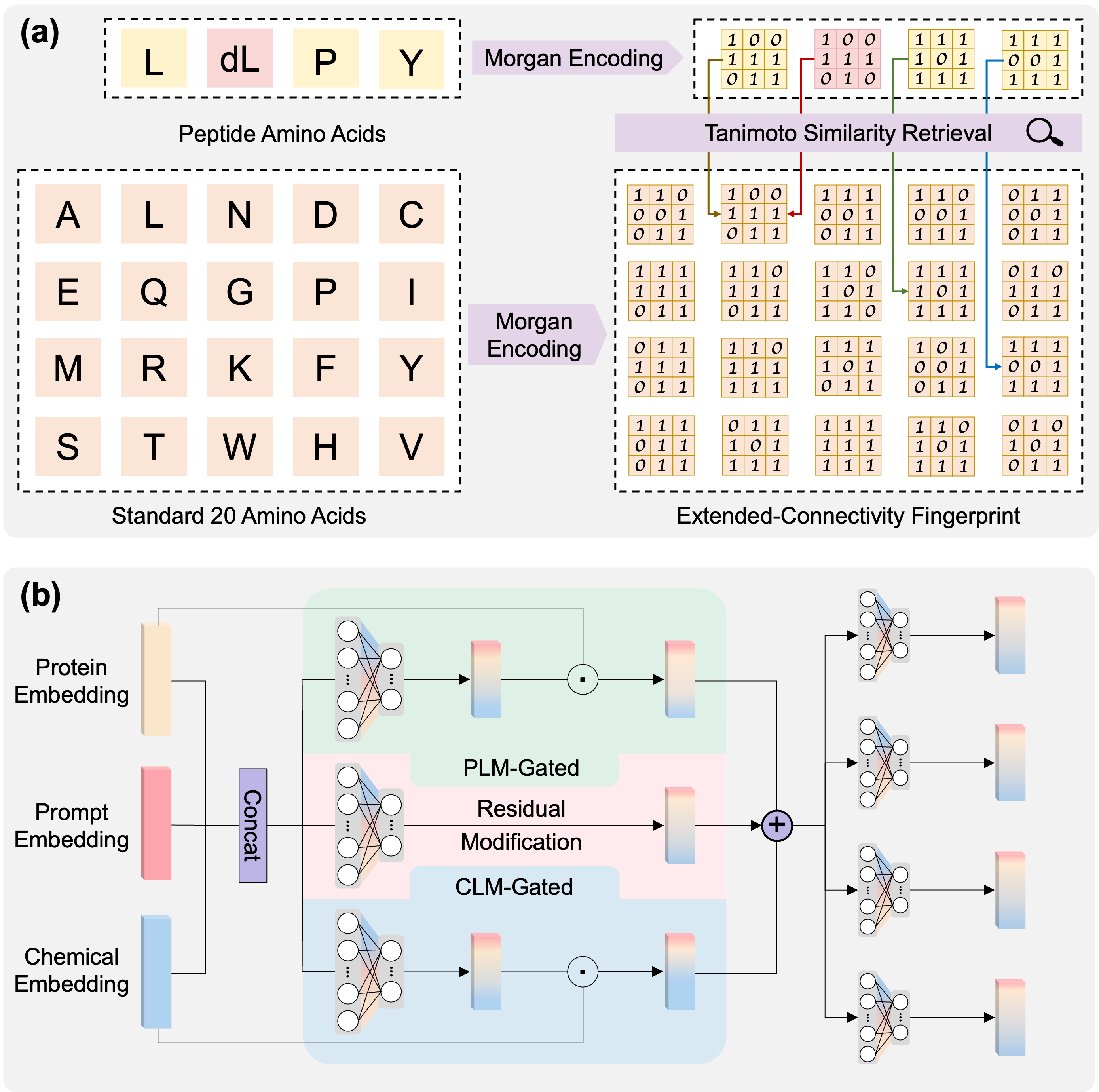}
    \caption{\textbf{Submodules of Bi-TEAM.} \textbf{(a)} Mapping workflow from non-canonical to natural amino acids. \textbf{(b)} The architecture of space modification via Bi-Gated residuAl fusIoN (Bi-GAIN) between biochemical representations and modification prompt position embedding. The PLM-Gated and CLM-Gated networks each use the biological and chemical domains as primary inputs, respectively, while selectively integrating signals from the other domain. Residual Modification utilizes residual layers to retrieve and integrate interdependent signals from both domains. The final output produces task-specific fusion representations for prediction.}
    \label{submodule}
\end{figure}

We employ a molecular similarity analysis that identifies the most closely related canonical amino acid for each ncAA, with the mapping process illustrated in Fig.~\ref{submodule}(a). Specifically, let $\mathbf{P}_{NA} = \{\mathbf{p}_{na}^1, \mathbf{p}_{na}^2, \ldots, \mathbf{p}_{na}^n\}$ denote the set of all ncAAs to be mapped. Using the RDKit library, each ncAA in $\mathbf{P}_{NA}$ is converted into its corresponding Morgan fingerprint: $\mathbf{E}_{NA} = \{\mathbf{e}_{na}^1, \mathbf{e}_{na}^2, \ldots, \mathbf{e}_{na}^n\}$. Morgan fingerprints capture topological features in a manner proven effective for similarity computation~\cite{ozturk2016comparative,galeano2016drug}. 
Similarly, we define the canonical amino acids: $\mathbf{P}_{CA} = \{\mathbf{p}_{ca}^1, \mathbf{p}_{ca}^2, \ldots, \mathbf{p}_{ca}^{20}\}$, and convert each of them into Morgan fingerprints under identical parameters: $\mathbf{E}_{CA} = \{\mathbf{e}_{ca}^1, \mathbf{e}_{ca}^2, \ldots, \mathbf{e}_{ca}^{20}\}$.

We next compute the Tanimoto similarity coefficient~\cite{vogt2020ccbmlib} between every $\mathbf{e}_{na}^i \in \mathbf{E}_{NA}$ and each $\mathbf{e}_{ca}^j \in \mathbf{E}_{CA}$. The Tanimoto similarity, defined as
\begin{equation}
\mathrm{Tanimoto}(\mathbf{e}_{na}^i, \mathbf{e}_{ca}^j) = 
\frac{\mathbf{e}_{na}^i \cdot \mathbf{e}_{ca}^j}{\|\mathbf{e}_{na}^i\|^2 + \|\mathbf{e}_{ca}^j\|^2 - \mathbf{e}_{na}^i \cdot \mathbf{e}_{ca}^j}
\end{equation}
ranges from 0 to 1, where higher values indicate closer structural resemblance. For each ncAA, we retain only the canonical amino acid with the highest similarity score that exceeds our predetermined threshold. If no sufficiently similar canonical residue is found, the ncAA is labeled as ''$\mathbf{X}$''. Formally, we denote the complete mapping process by Equation~\ref{eq:pa}.

This approach ensures that non-canonical residues are replaced by only the most structurally relevant canonical amino acids, preserving evolutionary signals and structural information gleaned from large protein corpora while avoiding the addition of underrepresented or anomalous tokens.
Consequently, the resulting biological embeddings, computed over these mapped residues, better align with learned protein-space features, facilitating downstream tasks such as property prediction for inputs containing non-canonical modifications.

\section{Bi-Gated Residual Fusion}\label{appendix:bi-gain}

We achieve space modification via a Bi-Gated ResiduAl FusIoN (Bi-GAIN) mechanism, which integrates the biological-space embedding $\mathbf{r}_p$, the chemical-space embedding $\mathbf{r}_c$, and the modification prompt embedding $\mathbf{r}_l$ into a unified representation, as illustrated in Fig.~\ref{submodule}(b). As outlined in the main text, this integration architecture comprises two distinct components: a bi-gated network and a residual network. Crucially, both submodules simultaneously incorporate the modification location prompt $\mathbf{r}_l$ to precisely pinpoint the position of non-canonical amino acids within the sequence. The final fused embedding is derived by summing the outputs of these two complementary pathways:
\begin{equation}
\mathbf{r}_{m} 
= \underbrace{f_{\mathrm{Bi\_Gated}}(\mathbf{r}_p,\mathbf{r}_c,\mathbf{r}_l)}_{\mathbf{r}_g}
\;+\;
\underbrace{f_{\mathrm{Res}}(\mathbf{r}_p,\mathbf{r}_c,\mathbf{r}_l)}_{\mathbf{r}_r}
\label{eq:bigated_fusion}
\end{equation}

\paragraph{Bi-Gated Network}
The bi-gated network employs a dual-stream strategy, treating each domain (biological or chemical) as a principal feature stream that is conditionally modulated by signals from the other. This approach preserves the inherent strengths of both representations: (i) \(\mathbf{r}_p\), which encodes broader sequence semantics and evolutionary cues, and (ii) \(\mathbf{r}_c\), which emphasizes fine-grained chemical details. Concretely, we learn two gating factors to modulate \(\mathbf{r}_p\) and \(\mathbf{r}_c\) based on the prompts \(\mathbf{r}_l\) and the combined features:
\begin{align}
\mathbf{r}_g = f_{\mathrm{Bi\_Gated}}(\mathbf{r}_p,\mathbf{r}_c,\mathbf{r}_l)
&=
\underbrace{\sigma\!\Bigl(\mathbf{W}_{g2}\,\operatorname{ReLU}\bigl(\mathbf{W}_{g1}\!\times [\mathbf{r}_p,\mathbf{r}_c,\mathbf{r}_l]\bigr)\Bigr)}_{\text{PLM Gated}}
\;\odot\;\mathbf{r}_p
\nonumber \\
&\quad+\;
\underbrace{\sigma\!\Bigl(\mathbf{W}_{g4}\,\operatorname{ReLU}\bigl(\mathbf{W}_{g3}\!\times [\mathbf{r}_p,\mathbf{r}_c,\mathbf{r}_l]\bigr)\Bigr)}_{\text{CLM Gated}}
\;\odot\;\mathbf{r}_c
\end{align}
where \([\mathbf{r}_p,\mathbf{r}_c,\mathbf{r}_l]\) denotes concatenation along the feature dimension, and \(\odot\) denotes elementwise multiplication. Through these gated activations, the model achieves a location-aware fusion: it retains the unique characteristics of the primary domain while injecting relevant complementary context, with the intensity of this fusion governed by the modification prompt.

\paragraph{Residual Network}

Operating in parallel, the residual network functions as a global interaction learner, designed to capture high-order non-linear synergies among the biological, chemical, and location embeddings:

\begin{align}
\mathbf{r}_r =
f_{\mathrm{Res}}(\mathbf{r}_p, \mathbf{r}_c, \mathbf{r}_l) =
\sigma\!\Bigl(
\mathbf{W}_{r2}\,\operatorname{ReLU}\bigl(
\mathbf{W}_{r1} \!\times [\mathbf{r}_p,\mathbf{r}_c,\mathbf{r}_l]
\bigr)
\Bigr)
\end{align}
Unlike the domain-specific gating, this sub-network models the holistic context, explicitly encoding how localized chemical modifications (ncAA) perturb the global peptide landscape. The final representation \(\mathbf{r}_{m}\) is constructed via the superposition of the gated features \(\mathbf{r}_g\) and the residual signals \(\mathbf{r}_r\). This additive fusion mechanism ensures the model creates a comprehensive embedding that balances domain-distinctive semantics with system-level dependencies, thereby facilitating robust property prediction.

\end{appendices}

\bibliography{sn-bibliography}% common bib file

\end{document}